\documentclass[letterpaper, 10 pt, conference]{ieeeconf}
\IEEEoverridecommandlockouts                             
\usepackage{amsfonts}
\usepackage{amsmath} 
\usepackage{amssymb}  

\usepackage{float}

\usepackage{hhline}
\usepackage{bm}
\usepackage{makecell}
\usepackage{multirow}
\newcolumntype{C}[1]{>{\centering\let\newline\\\arraybackslash\hspace{0pt}}m{#1}}

\usepackage{amsthm}

\theoremstyle{definition}

\usepackage{hyperref}
\hypersetup{
    colorlinks=true,
    linkcolor=blue,
    filecolor=magenta,      
    urlcolor=cyan,
}

\usepackage{tikz}
\usetikzlibrary{calc,patterns,decorations.pathmorphing,decorations.markings}
\usetikzlibrary{shapes,arrows}
\usepackage{verbatim}

\tikzstyle{block} = [draw, fill=blue!20, rectangle, 
    minimum height=3em, minimum width=6em]
\tikzstyle{sum} = [draw, fill=blue!20, circle, node distance=1cm]
\tikzstyle{input} = [coordinate]
\tikzstyle{output} = [coordinate]
\tikzstyle{pinstyle} = [pin edge={to-,thin,black}]

\usetikzlibrary{positioning}
\usetikzlibrary{math}

\usepackage{tikz}
\usetikzlibrary{shapes,arrows,calc,positioning}
\tikzstyle{bigblock} = [draw, fill=blue!20, rectangle, 
    minimum height=6em, minimum width=8em]
\tikzstyle{medblock} = [draw, fill=blue!20, rectangle, 
    minimum height=4em, minimum width=4em]    
\tikzstyle{mux} = [draw, fill=black!20, rectangle, 
    minimum height=5em, minimum width=0.1em]    
\tikzstyle{smallblock} = [draw, fill=blue!20, rectangle, 
    minimum height=3em, minimum width=4em]
\tikzstyle{sum} = [draw, fill=blue!20, circle, node distance=1cm]
\tikzstyle{signal} = [coordinate]
\tikzstyle{pinstyle} = [pin edge={to-,thin,black}]
\tikzstyle{block} = [draw, fill=blue!20, rectangle, 
    minimum height=3em, minimum width=6em]
\tikzstyle{blockS} = [draw, fill=blue!20, rectangle, 
    minimum height=3em, minimum width=4em]    
\tikzstyle{input} = [coordinate]
\tikzstyle{output} = [coordinate]
\usetikzlibrary{matrix}

\newcommand{\bc}{\begin{center}}
\newcommand{\ec}{\end{center}}
\newcommand{\benum}{\begin{enumerate}}
\newcommand{\eenum}{\end{enumerate}}
\newcommand{\nn}{\nonumber}
\newcommand{\matl}{\left[ \begin{array}}
\newcommand{\matr}{\end{array} \right]}
\newcommand{\matls}{\left[ \begin{smallmatrix}}
\newcommand{\matrs}{\end{smallmatrix} \right]}
\newcommand{\isdef}{\stackrel{\triangle}{=}}

\newcommand{\vect}[1]{\overset{\rightharpoonup}{#1}}

\newcommand{\rmE}{{\rm E}}

\newcommand{\rmG}{{\rm G}}

\newcommand{\rmI}{{\rm I}}

\newcommand{\rmN}{{\rm N}}

\newcommand{\rmT}{{\rm T}}

\newcommand{\rmW}{{\rm W}}

\newcommand{\rmc}{{\rm c}}
\newcommand{\rmd}{{\rm d}}

\newcommand{\rmm}{{\rm m}}

\newcommand{\rms}{{\rm s}}

\newcommand{\BBR}{{\mathbb R}}

\newcommand{\shiftq}{{\textbf{\textrm{q}}}}

\newcommand{\resolvedinFrame}[1]{{\big|_{\rm #1}}}

\newcommand{\ihat}{ {\hat \imath}}
\newcommand{\jhat}{ {\hat \jmath}}
\newcommand{\khat}{ {\hat k}}

\usepackage[style=ieee ,sorting=none,maxbibnames=99,giveninits, doi=false]{biblatex}
\title{An A Quadcopter Autopilot Based on an Adaptive Digital PID Controller}

\title{Adaptive Digital PID Control of a Quadcopter}

\title{Retrospective-Cost-Based Adaptive Digital PID Control of a Quadcopter}

\title{A Retrospective-Cost-Based Adaptive Digital PID   Quadcopter Autopilot}

\title{Adaptive Digital PID Control of a Quadcopter with Unknown Dynamics}

\title{One-Shot Learning for a Quadcopter Autopilot}

\title{An adaptive digital autopilot for Multicopters}

\title{\LARGE \bf Experimental Implementation of an Adaptive Digital Autopilot with Applications }

\title{\LARGE \bf An Adaptive Digital Autopilot with Applications \\ for Fixed-wing Aircraft Control }

\title{\LARGE \bf An Adaptive Digital Autopilot\\ for Fixed-Wing Aircraft with Actuator Faults}

\title{\LARGE \bf Experimental Flight Testing of a\\  Fault-Tolerant Adaptive Autopilot for Fixed-Wing Aircraft}

\author{
    Joonghyun Lee,
    John Spencer,
    Siyuan Shao,
    Juan Augusto Paredes, 
    Dennis S. Bernstein,
    Ankit Goel%
\thanks{This research was supported in part by the Office of Naval Research under grant N00014-19-1-2273.}
\thanks{Joonghyun Lee, John Spencer, Siyuan Shao, Juan Augusto Paredes, and Dennis S. Bernstein are with the Department of Aerospace Engineering, University of Michigan, Ann Arbor, MI 48109.
{\tt\small joonghle, spjohn, shaosy, jparedes,}
{\tt \small dsbaero@umich.edu}
}
\thanks{Ankit Goel is with the Department of Mechanical Engineering, University of Maryland, Baltimore County, MD 21250.
{\tt \small ankgoel@umbc.edu}}
}

\date{}

\begin{document}

\maketitle

\begin{abstract}
    This paper presents an adaptive autopilot for fixed-wing aircraft and compares its performance with a fixed-gain autopilot.
    The adaptive autopilot is constructed by augmenting the autopilot architecture with adaptive control laws that are updated using retrospective cost adaptive control.  
    In order to investigate the performance of the adaptive autopilot, the default gains of the fixed-gain autopilot are scaled to degrade its performance.
    This scenario provides a venue for determining the ability of the adaptive autopilot to compensate for the degraded fixed-gain autopilot.
    Next, the performance of the adaptive autopilot is examined under failure conditions by simulating a scenario where one of the control surfaces is assumed to be stuck at an unknown angle. 
    The adaptive autopilot is also tested in physical flight experiments under degraded-nominal conditions, and the resulting performance improvement is examined.  
    
\end{abstract}

\section{Introduction}

Autonomous flight control of an aircraft under rapidly changing conditions requires an autopilot that can control the aircraft in uncertain environments and without detailed models. 
An autopilot for a fixed-wing aircraft typically consists of a set of trim commands along with low-level controllers to follow intermediate commands.
The trim conditions for an aircraft can be computed by solving nonlinear algebraic equations for trim equilibria \cite{mcclamroch2011steady}, but a detailed model of the aircraft aerodynamics is required.
Moreover, for low-cost aircraft that are usually repaired or modified onsite, the true aerodynamic properties may be different from nominal aerodynamics. 
Consequently, a fixed-gain autopilot may not be able to maintain performance in a rapidly changing environment or under failure conditions such as damaged wings or faulty actuators. 
In this scenario, an adaptive autopilot may be able to compensate for the lost performance by updating the autopilot gains accordingly. 
With these motivations in mind, this paper explores the use of an in situ learning technique to modify the autopilot during the flight. 


Various adaptive control techniques have been investigated for    fixed-wing aircraft control
\cite{nguyen2006dynamics}. 
A sliding mode fault-tolerant tracking control scheme was used for control of a fixed-wing UAV under actuator saturation and state constraints in \cite{9262225, 9476716}. 
A backstepping algorithm was used  in \cite{hirano2019controller} to design a nonlinear flight controller for a fixed-wing UAV with thrust vectoring. 
An MRAC-based technique was used to augment the control system to improve the dynamic performance of a fixed-wing aircraft in \cite{9189264}. 
However, these techniques rely on the availability of a sufficiently detailed model for the control system synthesis.

In contrast, the present paper uses the retrospective cost adaptive control (RCAC) algorithm to learn the autopilot gains from the measured data in situ. 
RCAC is a digital adaptive control technique that is applicable to stabilization, command following, and disturbance rejection. 
Instead of relying on a model of the system, RCAC uses the past measured data and past applied input to recursively optimize the controller gains.
RCAC is described in \cite{rahmanCSM2017}, and its extension to digital PID control is given in \cite{rezaPID}.
The application of RCAC for a multicopter autopilot are described in \cite{goel_adaptive_pid_2021,quadtuner2021}.




The contribution of this paper is the development of an adaptive autopilot for fixed-wing aircraft, and a comparison of its performance with a well-tuned fixed-gain autopilot under nominal conditions, performance recovery of a degraded-nominal autopilot, and performance improvement under actuator failure. 
In particular, this paper presents the potential advantages of an adaptive autopilot by investigating two scenarios. 
In the first scenario, a well-tuned fixed-gain controller is degraded by scaling all of the gains by a small factor, and it is shown that the adaptive autopilot is able to compensate for the degraded gains by learning the necessary gains.
This scenario is investigated both in simulation and in physical flight experiments. 
In the second scenario, the aircraft is simulated with a faulty aileron, thus emulating an actuator failure condition, and it is shown, in simulation experiments, that the adaptive autopilot improves the trajectory-tracking performance.
%

%

The paper is organized as follows:
Section \ref{sec:notation} defines the notation used in this paper, %
Section \ref{sec:PX4_autopilot} reviews the autopilot architecture implemented in the PX4 flight stack,
%
%
Section \ref{sec:adaptiveAugmentation} presents the adaptive augmentation of autopilot, 
Section \ref{sec:flight_tests_sim} presents the simulation flight tests,
and Section \ref{sec:flight_tests_sim} presents the outdoor flight tests.
Finally,  Section \ref{sec:conclusions} concludes the paper with a summary and future research directions.

\section{Notation}
\label{sec:notation}
Let $\rm F_E$ denote an Earth-fixed frame such that $\khat_\rmE$ is aligned with the acceleration due to gravity $\vect g.$
Let $\rm F_{AC}$ denote an aircraft-fixed frame such that $\ihat_{\rm AC}$ is aligned with the fuselage, 
$\jhat_{\rm AC}$ is along the wing, and 
$\khat_{\rm AC}$ is chosen to complete the right-handed frame. 
Note that $\khat_{\rm AC}$ points vertically down. 
Next, let $\rmc$ denote the center of mass of the aircraft, and let $w$ be an point fixed on Earth.
The coordinates of the aircraft relative to $w$ in the Earth frame are denoted by $r \isdef \vect r_{\rmc/w}\resolvedinFrame{E} \in \BBR^3.$
The velocity of the aircraft relative to $w$ in the Earth frame is  $ v\isdef  \vect v_{\rmc/w/\rmE}\resolvedinFrame{E} \in \BBR^3.$ 
%
Let $\Psi$, $\Theta,$ and $\Phi$ denote the 3-2-1 azimuthal, elevation, and bank Euler angles of the aircraft. 
The angular velocity of $\rm F_{AC}$ relative to  $\rm F_E$ in the aircraft-fixed frame is given by $\omega \isdef \vect \omega_{{\rm AC} / \rmE}\resolvedinFrame{\rm AC} \in \BBR^3.$
The angular acceleration of $\rm F_{AC}$ relative to  $\rm F_E$ in the aircraft-fixed frame is given by $\alpha \isdef \vect \alpha_{{\rm AC} / \rmE}\resolvedinFrame{\rm AC} \in \BBR^3.$
%
The measurement of the variable $x$ is denoted by $x_\rmm$, and the setpoint for the variable $x$ is denoted by $x_\rms.$
Finally, let $e_3 \isdef \matl{c c c} 0 & 0 & 1 \matr^{\rmT}.$

The angles $\Psi,$ $\Theta,$ and $\Phi$ comprise a 3-2-1 sequence of Euler angles that parameterize the orientation of $\rm F_{AC}$ relative to $\rm F_E.$ 
The components of $\omega$ are the yaw rate, pitch rate, and roll rate, which are different from the azimuth rate, elevation rate, and bank rate. 
Hence, integrating the components of $\omega$ does not yield the azimuthal, elevation, and bank Euler angles.
%
%
%
In fact, the relation between the Euler-angle rates and the components of $\omega$ is given by \eqref{eq:e2omega} in the following section.  
%

%
\section{Flight Control Architecture}
\label{sec:PX4_autopilot}
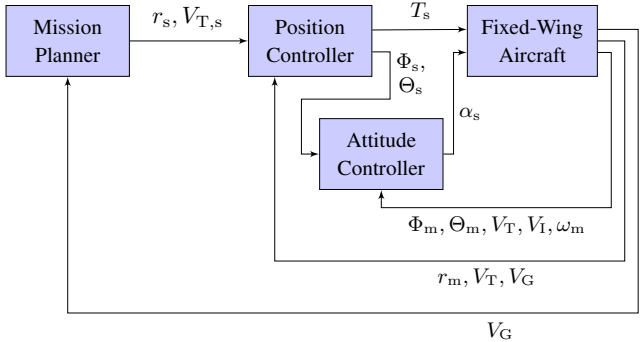
\begin{figure}[h]
    \centering
    \resizebox{\columnwidth}{!}{
    \begin{tikzpicture}[auto, node distance=2cm,>=latex',text centered]
    
        \node [smallblock, minimum height=3em, text width=1.6cm] (Mission) {\small Mission Planner};
        \node [smallblock, minimum height=3em, right = 5em of Mission, text width=1.6cm] (Pos_Cont) {\small Position Controller};
        \node [smallblock, minimum height=3em, below right = 1.75em and -2.25 em of Pos_Cont,text width=1.6cm] (Att_Cont) {\small Attitude Controller};
        \node [smallblock, minimum height=3em, text width=1.6cm, minimum width = 5.5em,  right = 4em of Pos_Cont] (FWAircraft) {\small Fixed-Wing Aircraft};
        
        \draw [->] (Mission) -- node[above, xshift = -0.05 em, yshift = 0.1em]{ {}$r_\rms, V_{\rmT, \rms}$} (Pos_Cont);
        \draw [->] (Pos_Cont.-10) -| ([xshift = 0.75em, yshift = -2em]Pos_Cont.-10) -| node  [xshift = 3.15em, yshift = 1em] {\small {}$\begin{array}{c} \Phi_{\rms} , \\ \Theta_{\rms} \end{array}$} ([xshift = -0.75em]Att_Cont.west) -- (Att_Cont.west);
        \draw [->] (Pos_Cont.10) -- node [above, xshift = 0.05em, yshift = 0.1 em] { \small{}$T _\rms$} (FWAircraft.170);
        \draw [->] (Att_Cont.0) -- +(0.15,0) |- node [below, xshift = 0.7 em, yshift = -2 em]{\small {}$\alpha_{\rms}$}(FWAircraft.190);
        \draw [->] (FWAircraft.0) -- +(.4,0) |- node[below,xshift = -6em, yshift = 0.05em]{ \small {} $r_{\rmm}, V_{\rmT}, V_{\rmG}$} ([xshift = -1.5em, yshift = -7.8 em]Pos_Cont.south) --  ([xshift = -1.5em]Pos_Cont.south);
        \draw [->] (FWAircraft.350) -- +(0.2,0) |- node[below,xshift = -5em]{ \small {} $ \Phi_{\rmm}, \Theta_{\rmm}, V_{\rmT}, V_{\rmI} , \omega_{\rmm}$} ([yshift = -0.75 em]Att_Cont.south) -- (Att_Cont.south);
         \draw [->] (FWAircraft.10) -- +(.6,0) |- node[below,xshift = -6em, yshift = -0.05em]{ \small {} $V_{\rmG}$} ([yshift = -10 em]Mission.south) --  (Mission.south);
    \end{tikzpicture}
    }
    \vspace{-2em}\caption{\footnotesize {Autopilot architecture.}}
    \label{fig:PX4_autopilot_nested_loop}
\end{figure}

In this work, we consider the flight control architecture implemented in the PX4 flight stack.
%
The control system consists of a mission planner and two cascaded controllers in nested loops as shown in Figure  \ref{fig:PX4_autopilot_nested_loop}.
The mission planner generates position setpoints based on user-defined waypoints. 

%

\begin{figure}[h]
    \vspace{0.5em}
    \centering
    \resizebox{1.0\columnwidth}{!}{
    \begin{tikzpicture}[auto, node distance=2cm,>=latex']
        \node[smallblock, minimum width = 6em, minimum height = 7 em] (TECS) 
        {$\begin{array}{c}
            {\rm Longitudinal} \\ 
            {\rm Controller} \\
            {\rm (TECS)}
        \end{array}$};
        %
        %
        \node[smallblock, below = 4.5em of TECS.center, minimum width = 2.5em, minimum height = 3 em] (L1C) {$\begin{array}{c}{\rm Lateral} \\ {\rm Controller}\end{array}$};
        \node[smallblock, below left = 0.05em and 7em of TECS.center, minimum width = 1em, minimum height = 1 em] (r2h1) {\small $e_3^\rmT$};
        \node[smallblock, below left = 2em and 7em of TECS.center, minimum width = 1em, minimum height = 1 em] (r2h2) {\small $e_3^\rmT$};
        \draw[->] (r2h1.east) -- node[above, xshift = -0.1em]{\small$h_{\rms}$} ([yshift = -0.9em]TECS.west);
        \draw[->] (r2h2.east) -- node[above]{\small$h_{\rmm}$} ([yshift = -2.875em]TECS.west);
        \draw[->] ([xshift = -4.5em]r2h1.west) -- node[above, xshift = -1.5em]{\small$r_\rms$} (r2h1.west);
        \draw[->] ([xshift = -4.5em]r2h2.west) -- node[above, xshift = -1.5em]{\small$r_\rmm$} (r2h2.west);
        \draw[->] ([xshift = -4em, yshift = 2.675em]TECS.west) -- node[above, xshift = -0.1em]{\small$V_{\rmT, \rms}$} ([yshift = 2.675em]TECS.west);
        \draw[->] ([xshift = -4em, yshift = 1em]TECS.west) -- node[above]{\small$V_{\rmT}$} ([yshift = 1em]TECS.west);
        \draw[->] ([xshift = -4em, yshift = 0.75em]L1C.west) -- node[above]{\small$V_{\rmG}$} ([yshift = 0.75em]L1C.west);
        \draw[-] ([xshift = -1.5em]r2h1.west)--([xshift = -1.5em, yshift = -1.75em]r2h1.west);
        \draw ([xshift = -1.5em, yshift = -2.25 em]r2h1.west) arc (270:90:0.25em);
        \draw[->] ([xshift = -1.5em, yshift = -2.25em]r2h1.west) |- (L1C.west);
        \draw[->] ([xshift = -2.5em]r2h2.west) |- ([yshift = -0.75em]L1C.west);
        \draw[->] ([yshift = 1.7em]TECS.east) -- node[above]{\small $T_{\rms}$} ([xshift = 3em, yshift = 1.7em]TECS.east);
        \draw[->] ([yshift = -1.7em]TECS.east) --  node[above]{\small $\Theta_{\rms}$} ([xshift = 3em, yshift = -1.7em]TECS.east);
        \draw[->] (L1C.east) --  node[above]{\small $\Phi_{\rms}$} ([xshift = 3em]L1C.east);
    \end{tikzpicture}
    }
    \caption{\footnotesize {Position controller architecture.}}
    \label{fig:PX4_autopilot_outer_loop}
\end{figure}
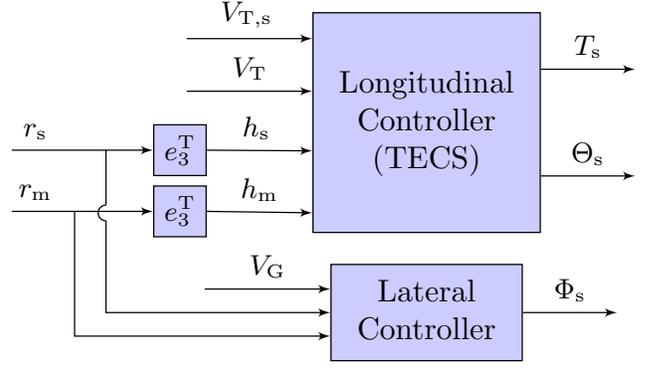

%
%
%
The outer loop, also called the position controller, consists of two decoupled controllers for the longitudinal and lateral motion of the aircraft, as shown in Figure \ref{fig:PX4_autopilot_outer_loop}. 
The longitudinal controller is based on the total energy control system (TECS) described in \cite{bruce1986,faleiro1999,lambregts2013,Argyle2016}, and the lateral controller is based on the guidance law described in \cite{park2004}.
%
The inputs to the position controller are the true airspeed setpoint $V_{\rmT, \rms},$ the position setpoint $r_\rms,$ the true airspeed $V_{\rmT},$ the position measurement $r_\rmm,$ and the ground velocity $V_{\rmG}.$ 
The TECS input includes the altitude setpoint $h_\rms \isdef e_3^\rmT r_\rms$ and the altitude measurement $h _\rmm \isdef e_3^\rmT r_\rmm.$
The longitudinal controller generates the thrust and the elevation setpoint, and the lateral controller generates the bank setpoint. 
%
%
The output of the position controller is thus the thrust setpoint $T_\rms$ and the attitude setpoint $\Psi_\rms, \Theta_\rms,\Phi_\rms.$

\begin{figure}[h]
    \centering
    \vspace{0.5em}
    \resizebox{\columnwidth}{!}{
    \begin{tikzpicture}[auto, node distance=2cm,>=latex']
    \node[smallblock, minimum width = 2.5em, minimum height = 4 em] (e2q) { \eqref{eq:e2omega}};
    \node[smallblock, above left = -0.5em and 4em of e2q.center, minimum width = 2.5em, minimum height = 2.5em] (G_E) {\eqref{eq:elevation_rate_P}, \eqref{eq:Roll_rate_P}};
    \node[smallblock, below left = 1.5em and 4.5em of e2q.center, minimum width = 2.5em, minimum height = 2.5em] (tc) {\eqref{eq:coordinated_turn}};
    \node[sum, left = 2.5em of G_E.center] (suml1){};
    \node[draw = none] at (suml1.center) {$+$};
    %
    %
    \node[sum, right = 2.5 em of e2q] (sumr1){};
    \node[draw = none] at (sumr1.center) {$+$};
    \node[smallblock, right = 1em of sumr1, minimum width = 2.5em, minimum height = 1.75 em] (G_PI) {$G_{\omega, {\rm PI}}$};
    \node[smallblock, above = 1em of G_PI, minimum width = 2.5em, minimum height = 1.75 em] (G_FF) {$G_{\omega, {\rm FF}}$};
    \node[smallblock, right = 1em of G_PI, minimum width = 2.5em, minimum height = 1.75 em] (sc) {\eqref{eq:alpha_s}};
    \draw [->] (sc.east) -- ([xshift = 1.5em]sc.east) node [above, xshift=-0.4 em] { {}$\alpha_s$};
    %
    \draw[->]([yshift = -0.25em]G_E.east) -- node[above]{ {} $\begin{bmatrix} \dot{\Phi}_{\rms} \\ \dot{\Theta}_{\rms} \end{bmatrix}$}([yshift = 0.5em]e2q.west);
    \draw[->](tc.east)-|([xshift = -2em, yshift = -1.25em]e2q.west)-- node[above, xshift = -0.35 em]{ {}$\dot{\Psi}_{\rms}$} ([yshift = -1.25em]e2q.west);
    %
    \draw[->](suml1.east) -- (G_E.west);
    \draw[->]([xshift = -0.75 em]suml1.west) |- ([yshift = 0.5em]tc.west);
    \draw[->]([xshift = -5.5em, yshift = -0.5em]tc.west) -- node[above, xshift = -2em]{ {}$V_{\rmm}$} ([yshift = -0.5em]tc.west);
    \draw[->]([xshift = -3.25em]suml1.west) -- node[above, xshift = -1em]{ {}$\begin{bmatrix}\Phi_{\rms} \\ \Theta_{\rms}\end{bmatrix}$} (suml1.west);
    \draw[->]([yshift = 1em]suml1.north) -- node[xshift = -1.4 em, yshift = 1.75em]{ {}$\begin{bmatrix}\Phi_{\rmm} \\ \Theta_{\rmm} \end{bmatrix}$} node[xshift = -0.15em, yshift = -0.05em]{$-$} (suml1.north);
    \draw[->](e2q.east)--node[above, xshift = -0.4em]{ {}$\omega_{\rms}$}(sumr1.west);
    \draw[->](sumr1.east) -- (G_PI.west);
    \draw[->]([yshift = -1em]sumr1.south) -- node[xshift = 1em, yshift = -1em]{ {}$\omega_{\rmm}$} node [xshift = 1.25em, yshift = -0.1em]{$-$} (sumr1.south);
    \draw[->]([xshift = 1.75em]e2q.east) |- (G_FF.west);
    \draw[->](G_FF.east)-|
    (sc.north);
    \draw[->](G_PI.east) -- 
    (sc.west);
    \draw[->]([yshift = -1em]sc.south) -- node[xshift = 1.75em, yshift = -1.25em]{ {}$V_{\rmT}, V_{\rmI}$} (sc.south);
    \end{tikzpicture}
    }
    \caption{\footnotesize {Attitude controller architecture.}}
    \label{fig:PX4_autopilot_inner_loop}
\end{figure}
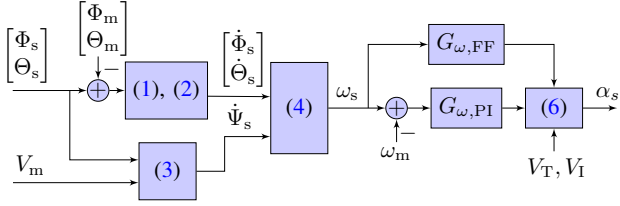

The inner loop, also called the attitude controller, consists of two cascaded controllers, as shown in Figure \ref{fig:PX4_autopilot_inner_loop}. 
The first controller uses the elevation and bank errors and a proportional control law to generate the elevation-rate and bank-rate setpoints. 
In particular, the elevation-rate setpoint $\dot{\Theta}_{\rms}$ and the bank-rate setpoint $\dot{\Phi}_{\rms} $ are given by
\begin{align}
    \dot{\Theta}_{\rms} 
        &=
            k_\theta(\Theta_{\rms} - \Theta_{\rmm})
    \label{eq:elevation_rate_P}
        , \\        
    \dot{\Phi}_{\rms} 
        &=
            k_\phi (\Phi_{\rms} - \Phi_{\rmm}), 
    \label{eq:Roll_rate_P}
\end{align}
where $k_{\theta}, k_{\phi}$ are the scalar gains.
The azimuthal-rate is algebraically given by 
\begin{align}
    \dot{\Psi}_{\rms} 
        = \frac{g \tan\Phi_{\rms} \cos\Theta_{\rms}}{V_{\rmT}}
    \label{eq:coordinated_turn}
\end{align}
to ensure coordinated turn. 
Finally, the body-fixed angular-velocity setpoint $\omega_\rms
$ is given by
\begin{align}
    \omega_\rms
        &= 
            S(\Theta_{\rmm}, \Phi_{\rmm})
            \begin{bmatrix} 
                \dot{\Phi}_{\rms} \\ \dot{\Theta}_{\rms} \\ \dot{\Psi}_{\rms} 
            \end{bmatrix},
    \label{eq:e2omega}
\end{align}
where
\begin{align}
    S(\Theta_{\rmm}, \Phi_{\rmm})
        \isdef
    \begin{bmatrix} 
        1 & 0 & \sin \Theta_{\rmm} \\
        0 & \cos \Phi_{\rmm} & \sin \Phi_{\rmm} \cos \Theta_{\rmm} \\
        0 & -\sin\Phi_{\rmm} & \cos\Phi_{\rmm} \cos\Theta_{\rmm}
    \end{bmatrix}.
\end{align}
Next, a feedforward and a PI control law generates the angular-acceleration setpoint $\alpha_s
$.
In particular, $\alpha_s $ is given by
\begin{align}
    \alpha_s
        &=
            \frac{V_{\rmT, 0}}{V_{\rmT}} G_{\omega,{\rm ff}} 
            \omega_\rms+
            \left(\frac{V_{\rmI,0}}{V_{\rmI}}\right)^2 G_{\omega,{\rm PI}}(\shiftq) 
            \left( \omega_\rms - \omega_\rmm
            \right), \label{eq:alpha_s}
\end{align}
where $G_{\omega,{\rm ff}} = k_{\omega, \rm ff} $ is a proportional control law, 
$G_{\omega,{\rm PI}}(\shiftq)  = k_{\omega, {\rm P}} + \dfrac{k_{\omega, {\rm I}}}{\textbf{q}-1}$ is a PI control law,
$V_{\rmI}$ is the indicated airspeed,
and $V_{\rmT,0}$ and $V_{\rmI,0}$ are the true airspeed and the indicated airspeed at trim conditions respectively, which are aircraft parameters.
%
Note that 
$\shiftq$ is the forward-shift operator, 
$k_{\omega, \rm ff}, $
$k_{\omega, \rm P},$ and
$k_{\omega, \rm I}$ are $3\times 3$ diagonal matrices, and are thus parameterized by 9 scalar gains. 
Finally, using the angular-acceleration setpoint, the actuator deflections are computed using control allocation methods.

The fixed-wing autopilot thus consists of 11 gains.
%
In practice, these 11 gains are tuned manually, which requires considerable expertise. 
We assume that the default gains implemented in PX4 are well tuned, and thus we refer to the autopilot with the default PX4 gains as the \textit{nominal autopilot.}

To investigate potential improvements and demonstrate the ability of the adaptive autopilot to recover performance, the gains in the nominal autopilot are multiplied by a scalar $\alpha_\rmd$ in order to degrade its performance.
A fixed-gain autopilot with the degradation factor $\alpha_\rmd \neq 1$ is referred to as the \textit{degraded-nominal autopilot}. 
Note that $\alpha_\rmd \neq 1$ is equivalent to the case of a poor choice of controller gains in the fixed-gain autopilot. 

%



\section{Adaptive Autopilot}
\label{sec:adaptiveAugmentation}
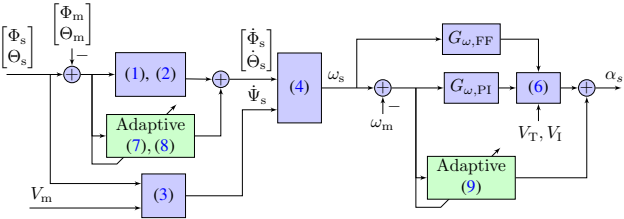
\begin{figure}[h]
    \centering
    \resizebox{\columnwidth}{!}{
    \begin{tikzpicture}[auto, node distance=2cm,>=latex']
    \node[smallblock, minimum width = 2.5em, minimum height = 4 em] (e2q) { \eqref{eq:e2omega}};
    \node[smallblock, above left = -0.5em and 6.5em of e2q.center, minimum width = 4em, minimum height = 2.5em] (G_E) {\eqref{eq:elevation_rate_P}, \eqref{eq:Roll_rate_P}};
    \node[smallblock, below left = 4.9em and 6.5em of e2q.center, minimum width = 2.5em, minimum height = 2.5em] (tc) {\eqref{eq:coordinated_turn}};
    \node[sum, left = 4em of G_E.center] (suml1){};
    \node[draw = none] at (suml1.center) {$+$};
    \node[sum, below right = -0.075em and 1.65em of G_E.east] (suml2){};
    \node[draw = none] at (suml2.center) {$+$};
    \node[sum, right = 3em of e2q] (sumr1){};
    \node[draw = none] at (sumr1.center) {$+$};
    \node [right = 0.3 em of suml1.east] (suml1_hook) {};
    \draw [->] (suml1_hook.center) -- +(0,-1.8) -- +(.5,-1.8) -- +(1.7,-.6);
    \node[smallblock, fill=green!20, below = 1em of G_E.south, minimum width = 0.5em, minimum height = 2 em, inner sep=0.25pt](G_E_adap){ {}$\begin{array}{c} {\rm Adaptive} \\ \eqref{eq:adaptive_elevation_rate_P}, \eqref{eq:adaptive_roll_rate_P} \end{array}$};
    \draw[->] (suml1_hook.center)|-(G_E_adap.west);
    \node[smallblock, right = 3em of sumr1, minimum width = 2.5em, minimum height = 1.75 em] (G_PI) {$G_{\omega, {\rm PI}}$};
    \node [left = 1.25 em of G_PI.west] (G_PI_hook) {};
    \draw [->] (G_PI_hook.center) -- +(0,-2.4) -- +(.5,-2.4) -- +(1.7,-1.2);
    \node[smallblock, fill=green!20, below = 3em of G_PI.south, minimum width = 0.5em, minimum height = 2 em, inner sep=0.25pt](G_PI_adap){ {}$\begin{array}{c} {\rm Adaptive} \\ \eqref{eq:alpha_s_adaptive} \end{array}$};
    \draw[->] (G_PI_hook.center)|-(G_PI_adap.west);
    \node[smallblock, above = 1em of G_PI, minimum width = 2.5em, minimum height = 1.75 em] (G_FF) {$G_{\omega, {\rm FF}}$};
    %
    %
    \node[smallblock, right = 1em of G_PI, minimum width = 2.5em, minimum height = 1.75 em] (sc) {\eqref{eq:alpha_s}};
    \node[sum, right = 1em of sc] (sumPI){};
    \node[draw = none] at (sumPI.center) {$+$};
    \draw [->] (sumPI.east) -- ([xshift = 1.5em]sumPI.east) node [above, xshift=-0.4 em] { {}$\alpha_s$};
    \draw[->]([yshift = -0.25em]G_E.east) -- (suml2.west);
    \draw[->](suml2.east) -- node[above]{ {} $\begin{bmatrix} \dot{\Phi}_{\rms} \\ \dot{\Theta}_{\rms} \end{bmatrix}$}([yshift = 0.5em]e2q.west);
    \draw[->](G_E_adap.east) -| (suml2.south);
    \draw[->](tc.east)-|([xshift = -2em, yshift = -1.25em]e2q.west)-- node[above, xshift = -0.15 em]{ {}$\dot{\Psi}_{\rms}$} ([yshift = -1.25em]e2q.west);
    \draw[->](suml1.east) -- (G_E.west);
    \draw[->]([xshift = -0.75 em]suml1.west) |- ([yshift = 0.7em]tc.west);
    \draw[->]([xshift = -6.4em, yshift = -0.7em]tc.west) -- node[above, xshift = -2.5em]{ {}$V_{\rmm}$} ([yshift = -0.7em]tc.west);
    \draw[->]([xshift = -3.25em]suml1.west) -- node[above, xshift = -1em]{ {}$\begin{bmatrix}\Phi_{\rms} \\ \Theta_{\rms}\end{bmatrix}$} (suml1.west);
    \draw[->]([yshift = 1em]suml1.north) -- node[xshift = -1.6 em, yshift = 1.9em]{ {}$\begin{bmatrix}\Phi_{\rmm} \\ \Theta_{\rmm} \end{bmatrix}$} node[xshift = -0.15em, yshift = 0.1em]{$-$} (suml1.north);
    \draw[->](e2q.east)--node[above, xshift = -0.65em]{ {}$\omega_{\rms}$}(sumr1.west);
    \draw[->](sumr1.east) -- (G_PI.west);
    \draw[->]([yshift = -1em]sumr1.south) -- node[xshift = 1em, yshift = -1.1em]{ {}$\omega_{\rmm}$} node [xshift = 1.4em, yshift = -0.1em]{$-$} (sumr1.south);
    \draw[->]([xshift = 2em]e2q.east) |- (G_FF.west);
    %
    %
    \draw[->](G_FF.east)-|
    (sc.north);
    \draw[->](G_PI_adap)-|(sumPI.south);
    \draw[->](G_PI.east)--(sc.west);
    \draw[->](sc.east) -- 
    (sumPI.west);
    \draw[->]([yshift = -1em]sc.south) -- node[xshift = 1.75em, yshift = -1.25em]{ {}$V_{\rmT}, V_{\rmI}$} (sc.south);
    \end{tikzpicture}
    }
    \caption{\footnotesize {Adaptive augmentation in the attitude controller.}}
    \label{fig:Augmented_PX4_autopilot_inner_loop}
\end{figure}

This section describes the adaptive autopilot, which is constructed by augmenting the nominal autopilot.
The nominal autopilot is the autopilot described in Section \ref{sec:PX4_autopilot} with fixed gains. 
In the adaptive autopilot, the fixed-gain control laws of the nominal autopilot are augmented with adaptive control laws, whose coefficients are updated by the retrospective cost adaptive control (RCAC) algorithm described in \cite{rahmanCSM2017,rezaPID}.
RCAC is used to augment the fixed-gain controllers of a multicopter autopilot in \cite{goel_adaptive_pid_2021,spencer2022adaptive}.
The output of a modified controller in the adaptive autopilot is thus given by the sum of the fixed-gain and the adaptive control law, as shown in Figure \ref{fig:Augmented_PX4_autopilot_inner_loop}. 

The bank and elevation rate setpoints $\dot{\Phi}_{\rms}, \dot{\Theta}_{\rms}$ in the adaptive autopilot are given by
%
\begin{align}
    \dot{\Theta}_{\rms} 
        &=
            k_\theta(\Theta_{\rms} - \Theta_{\rmm})
            + 
            u_\Theta
        \label{eq:adaptive_elevation_rate_P}
        , \\        
    \dot{\Phi}_{\rms} 
        &=
            k_\phi (\Phi_{\rms} - \Phi_{\rmm})
            +
            u_\Phi
            , 
        \label{eq:adaptive_roll_rate_P}
\end{align}
where the scalar adaptive control signals $u_\Theta$ and $u_\Phi$ are computed by RCAC.
Similarly, the angular acceleration setpoint in the adaptive autopilot is given by 
\small
\begin{align}
    \alpha_s
        &=
            \frac{V_{\rmT, 0}}{V_{\rmT}} G_{\omega,{\rm ff}} 
            \omega_\rms 
            \nn \\ 
            & \quad +
            \left(\frac{V_{\rmI,0}}{V_{\rmI}}\right)^2 G_{\omega,{\rm PI}} (\shiftq) 
            \left( \omega_\rms - \omega_\rmm
            \right)
            +
            u_{\omega, \rm PI}, \label{eq:alpha_s_adaptive}
\end{align}
\normalsize
where
$u_{\omega, \rm PI}$ is computed by RCAC.
Note that $ u_{\omega, \rm PI} \in \BBR^3,$ and each component of $u_{\omega, \rm PI}$ is updated by RCAC, where the error variable is the corresponding error term.


\section{Simulation Results}
\label{sec:flight_tests_sim}

In this section, we investigate the performance of the adaptive autopilot and compare it to the performance of the nominal autopilot, implemented in PX4, in the Gazebo simulation environment.
%
%
The aircraft dynamics simulated in Gazebo are based on the standard catapult-launched plane model\footnote{\href{https://docs.px4.io/main/en/simulation/gazebo_vehicles.html}{https://docs.px4.io/main/en/simulation/gazebo\_vehicles.html}} 
and are integrated in the PX4 version V1.13.0dev\footnote{\href{https://github.com/JAParedes/PX4-Autopilot/tree/RCAC\_FW\_UM}{https://github.com/JAParedes/PX4-Autopilot/tree/RCAC\_FW\_UM}}.
We also consider the case of a faulty actuator. 
To simulate a fault scenario, we assume that one of the ailerons is frozen at an unknown angle. 

Numerical simulations show that the aircraft performance is robust to TECS and the lateral controller gains.
Therefore, in this work, we focus on augmenting only the attitude controller with the adaptive control law, and thus $\alpha_\rmd$ degrades only the attitude controller in the nominal autopilot.
The hyperparameters $P_0,$ $R_u,$ and $\sigma$ used in RCAC are shown in Table \ref{tab:RCPE_variables_SITL}. 
Furthermore, we set $R_z=1$ in all adaptive controllers and all tests. 
Note that once the RCAC hyperparameters are tuned, they are fixed and thus they are not changed as $\alpha_\rmd$ is varied across the simulation tests. 

\begin{table}[h]
    \caption{\footnotesize RCAC hyperparameters used by the adaptive autopilot for all simulations. }
    \label{tab:RCPE_variables_SITL}
    \centering
    \renewcommand{\arraystretch}{1.2}
    \begin{tabular}{|c|l|l|l|}
        \hline
        \multicolumn{1}{|c|}{\textbf{Controller}}  & \multicolumn{1}{|c|}{${P_0}$}  & \multicolumn{1}{|c|}{${R_u}$} & \multicolumn{1}{|c|}{$\sigma$}
        \\ \hhline{|=|=|=|=|}
        \eqref{eq:adaptive_elevation_rate_P}, $\theta_\Theta$ & $1$ & $0.001$ & $-0.1$
        \\ \hline
        \eqref{eq:adaptive_roll_rate_P}, $\theta_\Phi$ & $1$ & $0.001$ & $-0.1$
        \\ \hline
    \end{tabular}
\end{table}

\begin{figure}[h]
    \centering
    \includegraphics[width=0.92\columnwidth]{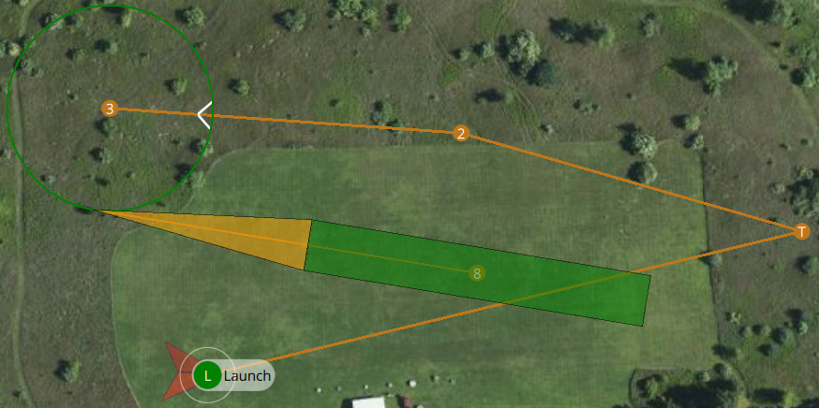}
    \caption{
    Waypoints used to construct the flight trajectory in simulation experiments.
    }
    \label{fig:FW_sim_trajectory}
\end{figure}

The mission waypoints are shown in Figure \ref{fig:FW_sim_trajectory}. 
The aircraft is assumed to be launched by a catapult from the launch point, and is commanded to fly toward the point $\rmT$ while climbing to an altitude of 20 m.
%
The aircraft is then commanded to fly around point 2 in a steady-state circular flight with a radius of 30 m for one minute. 
Finally, the aircraft is commanded to land along the green strip.


In order to quantify and compare the performance of the autopilot, bank, elevation, and trajectory-tracking error metrics are defined as
\begin{align}
    J_{\Phi} &\isdef \sqrt{\frac{1}{N} \sum_{i=1}^N (\Phi_{{\rms},i} - \Phi_{{\rmm},i})^2}, \label{eq:cost_bank} \\
    J_{\Theta} &\isdef \sqrt{\frac{1}{N} \sum_{i=1}^N (\Theta_{{\rms},i} - \Theta_{{\rmm},i})^2},\label{eq:cost_elevation} \\
    J_{\rm traj} &\isdef \sqrt{\frac{1}{N} \sum_{i=1}^N e_{{\rm x-track},i}^2},\label{eq:cost_xtrack}
\end{align}
where 
$N$ is the number of measurements during the flight, 
$e_{{\rm x-track}}$ is the cross-track error, which is defined as the minimum distance between the current position and desired trajectory. 
%
These error metrics are computed offline.

\begin{figure}[h!]
    \centering
    \includegraphics[width=\columnwidth]{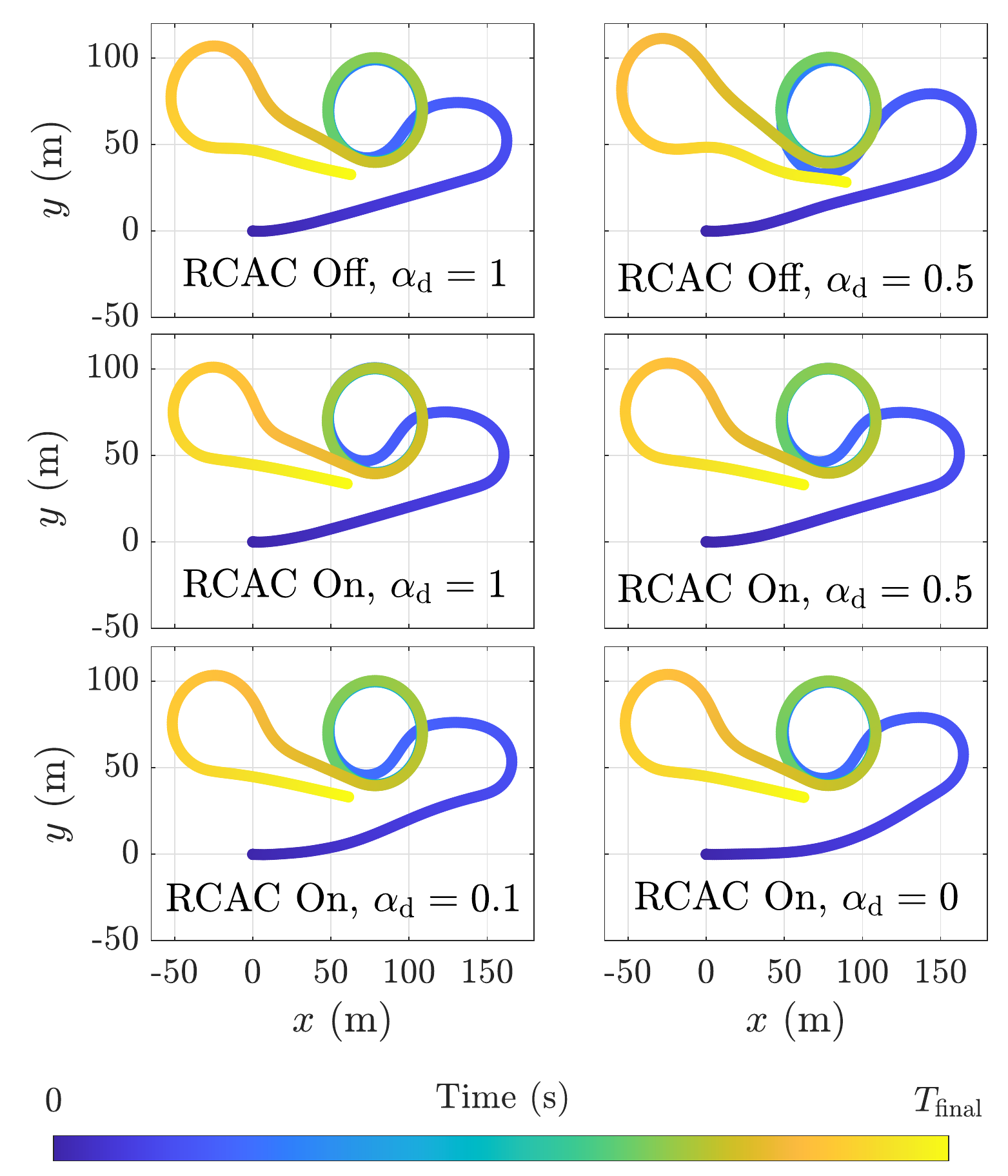}
    \caption{
    \textbf{Flight simulation.}
    Ground trace of the aircraft with the nominal, degraded-nominal, and the adaptive autopilot for several values of the degradation factor $\alpha_\rmd.$
    }
    \label{fig:Sim_comp_traj}
\end{figure}

\begin{figure}[h!]
    \centering
    \includegraphics[width=\columnwidth]{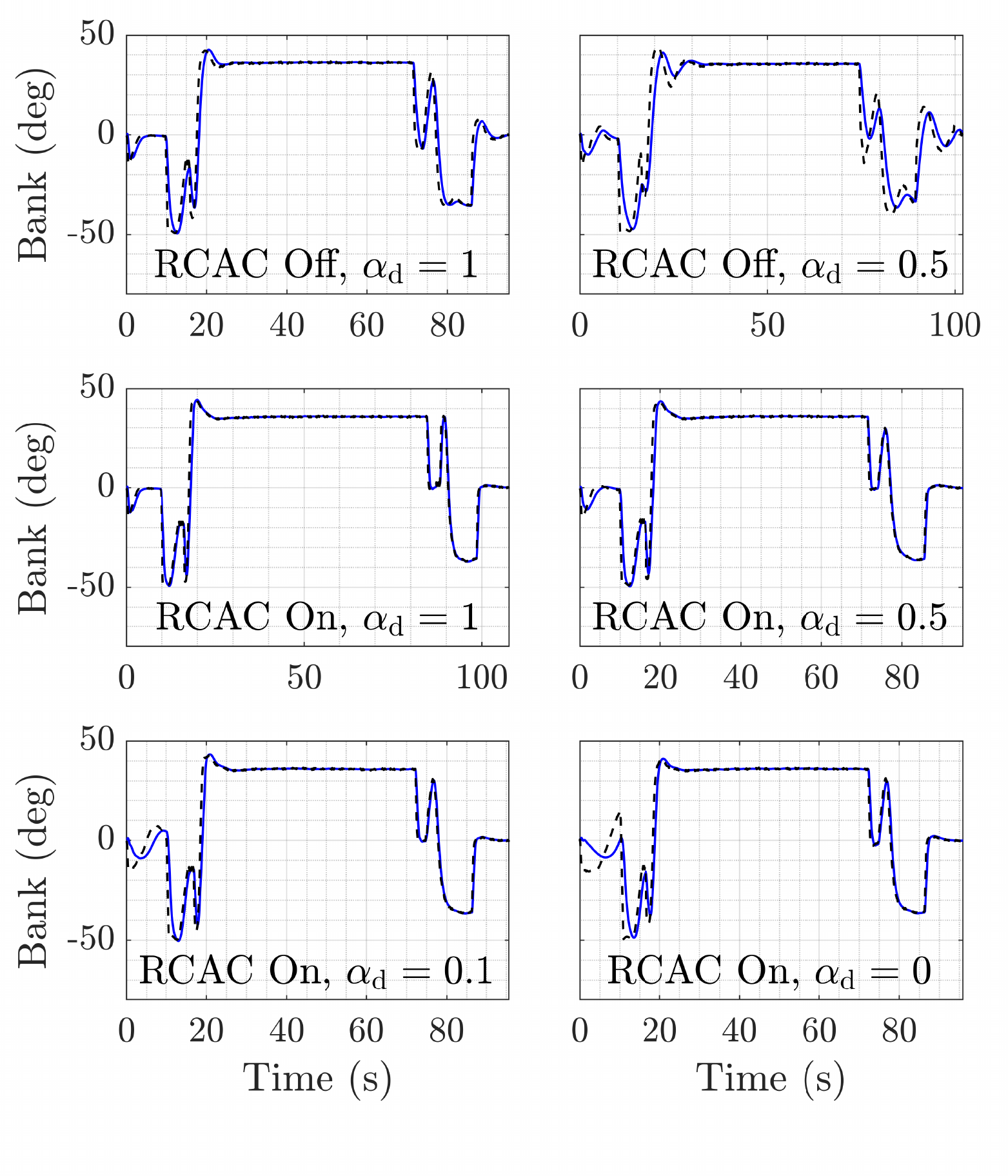}
    \vspace{-3.5em}
    \caption{
    \textbf{Flight simulation.}
    Bank-angle response of the aircraft with the nominal, degraded-nominal, and the adaptive autopilot for several values of the degradation factor $\alpha_\rmd.$
    The bank angle setpoints are displayed in black dashes.
    }
    \label{fig:Sim_comp_bank}
\end{figure}

\begin{figure}[h!]
    \centering
    \includegraphics[width=\columnwidth]{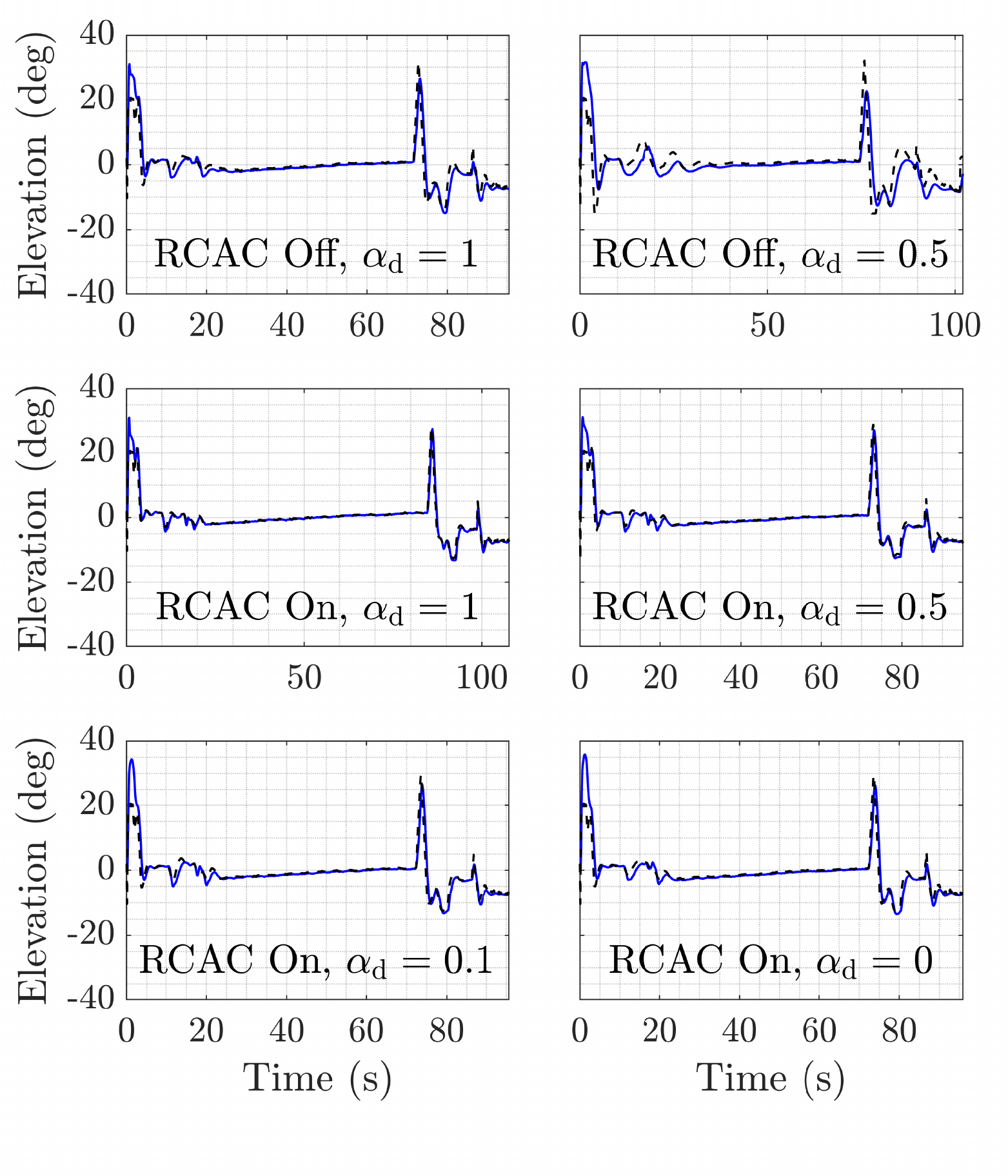}
    \vspace{-3.5em}
   \caption{
   \textbf{Flight simulation.}
   Elevation-angle response of the aircraft with the nominal, degraded-nominal, and the adaptive autopilot for several values of the degradation factor $\alpha_\rmd.$
   %
    %
    The elevation angle setpoints are indicated by black dashes. 
    }
    \vspace{-1.5em}
    \label{fig:Sim_comp_elevation}
\end{figure}

\begin{figure}[h!]
    \centering
    \includegraphics[width=\columnwidth]{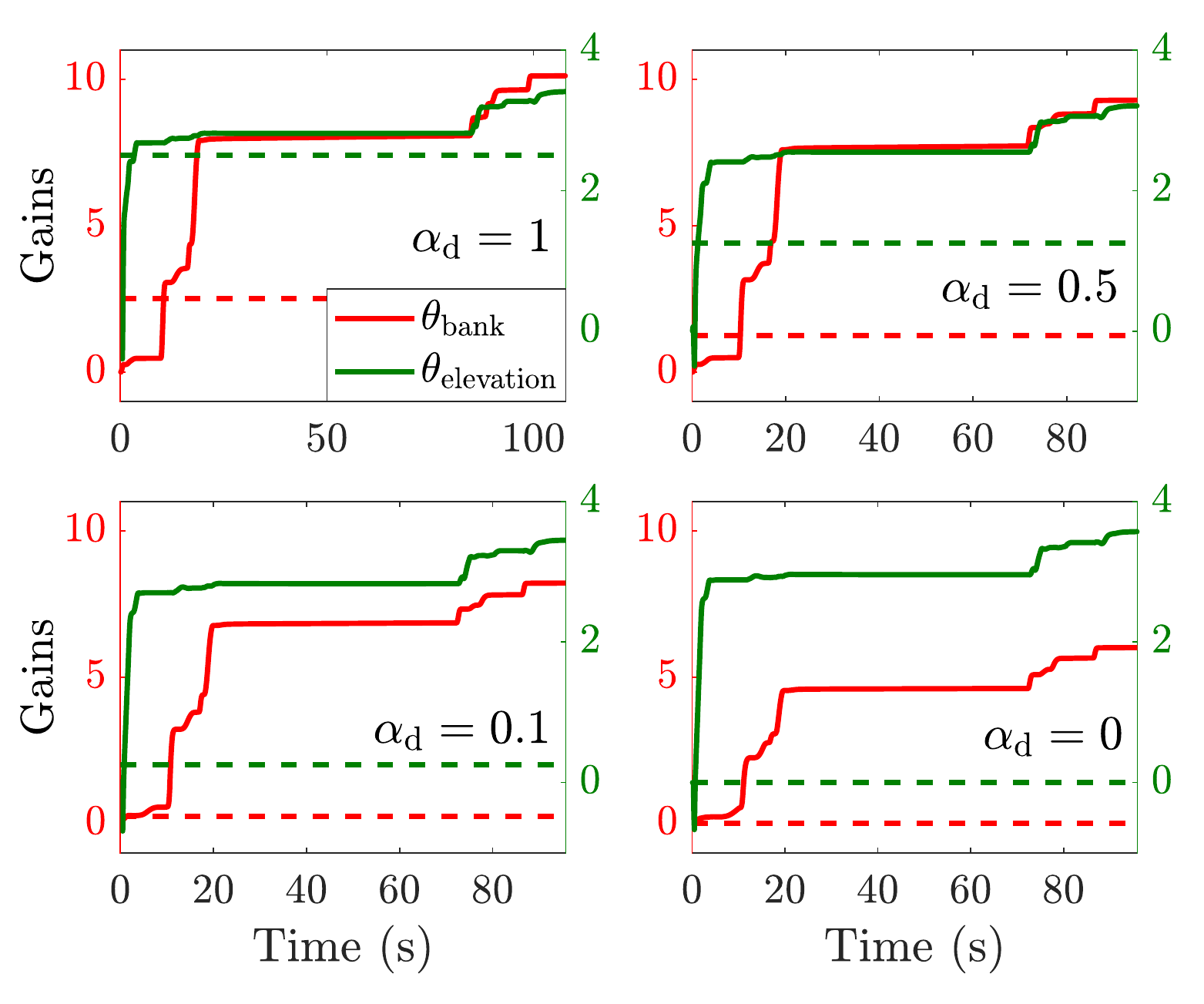}
    \vspace{-2em}
    \caption{
    \textbf{Flight simulation.}
    Adaptive bank and elevation controller gains optimized by RCAC in the adaptive autopilot for several values of the degradation factor $\alpha_\rmd.$
    In each case, the fixed gains are degraded by $\alpha_\rmd$ and RCAC updates the adaptive control laws.
    %
    %
    %
    %
    %
    The fixed gains are shown in dashes, and the adaptive gains are shown in solid for both the bank and elevator controller. 
    }
    \label{fig:Sim_comp_RCAC}
\end{figure}

\begin{figure}[h!]
    \centering
    \includegraphics[width=\columnwidth]{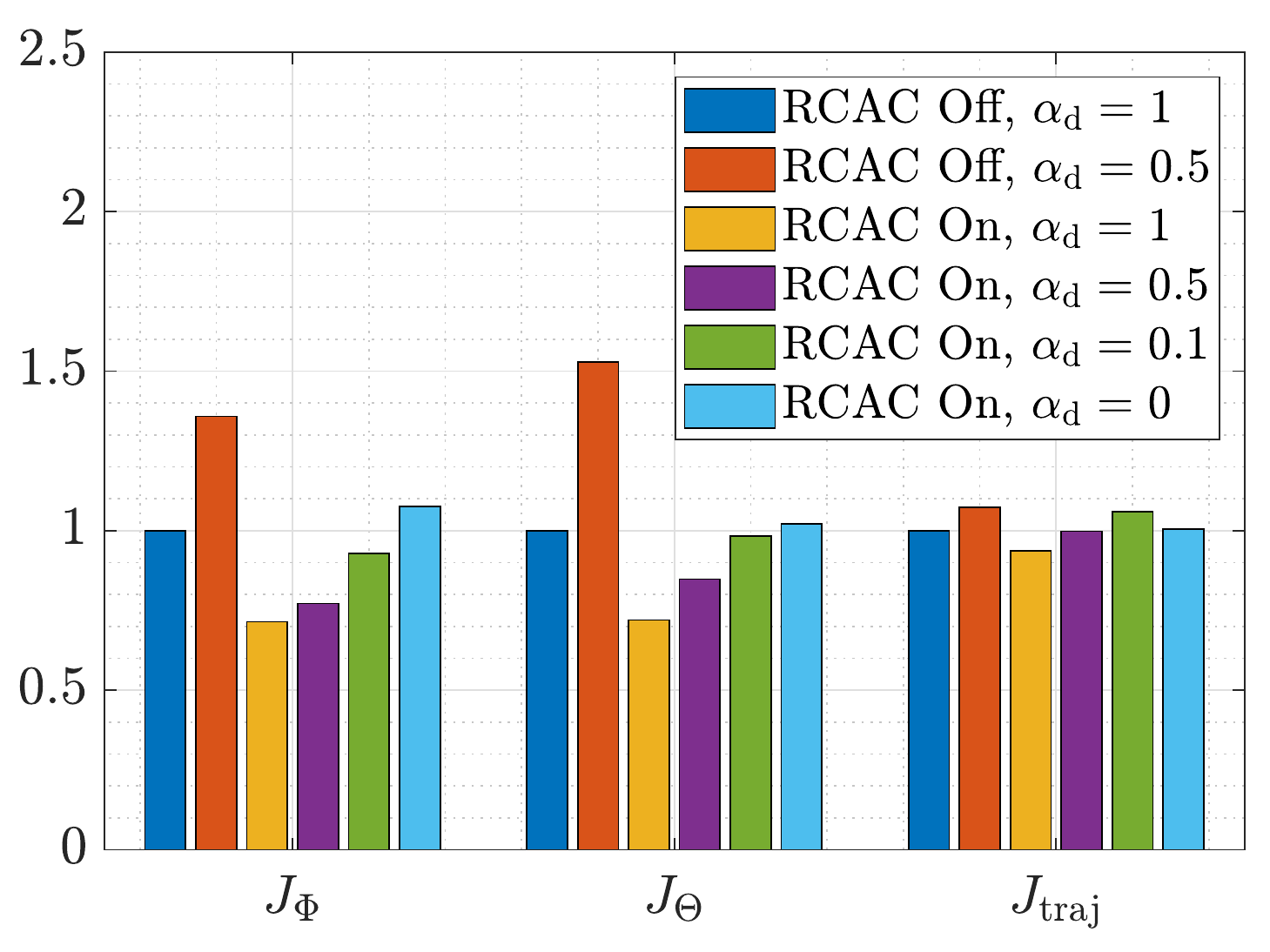}
    \vspace{-1.75em}
    \caption{
    \textbf{Flight simulation.}
    {Bank, elevation, and trajectory-tracking error metrics obtained with the 
    nominal, degraded-nominal, and  
    adaptive autopilots for several values of $\alpha_\rmd$}.
    Note that all metrics are normalized by the corresponding error metric obtained with the nominal fixed-gain autopilot.
    %
    }
    \vspace{-1em}
    \label{fig:Sim_J_cost}
\end{figure}

Figure \ref{fig:Sim_comp_traj} shows the ground trace,
Figure \ref{fig:Sim_comp_bank} shows the bank-angle response, and 
Figure \ref{fig:Sim_comp_elevation} shows the elevation-angle response of the aircraft with the nominal autopilot and the adaptive autopilot for several values of the degradation factor $\alpha_\rmd.$
Figure \ref{fig:Sim_comp_RCAC} shows the adaptive bank and elevation controller gains optimized by RCAC in the adaptive autopilot for several values of the degradation factor $\alpha_\rmd.$
Figure \ref{fig:Sim_J_cost} shows normalized error metrics for several values of $\alpha_\rmd$ with the
nominal, degraded-nominal, and adaptive autopilots.
%
%
The error metrics are normalized by the error metrics obtained with the nominal autopilot.
%

As shown in Figure \ref{fig:Sim_J_cost}, the adaptive autopilot improves the performance over the nominal performance. 
%
%
For $\alpha_\rmd=0.5,$ the trajectory following response degrades substantially with the degraded-nominal autopilot. 
In this case, the adaptive autopilot recovers the baseline performance. 
In fact, as the nominal controller is degraded, RCAC compensates by providing larger values of the corresponding gains. 
Finally, the adaptive autopilot is also able to learn the gains from a \textit{cold start}, that is, the case where the nominal autopilot is completely switched off, that is, $\alpha_\rmd=0.$

\begin{figure}[h!]
    \centering
    \vspace{-0.5em}
    \includegraphics[width=0.8\columnwidth]{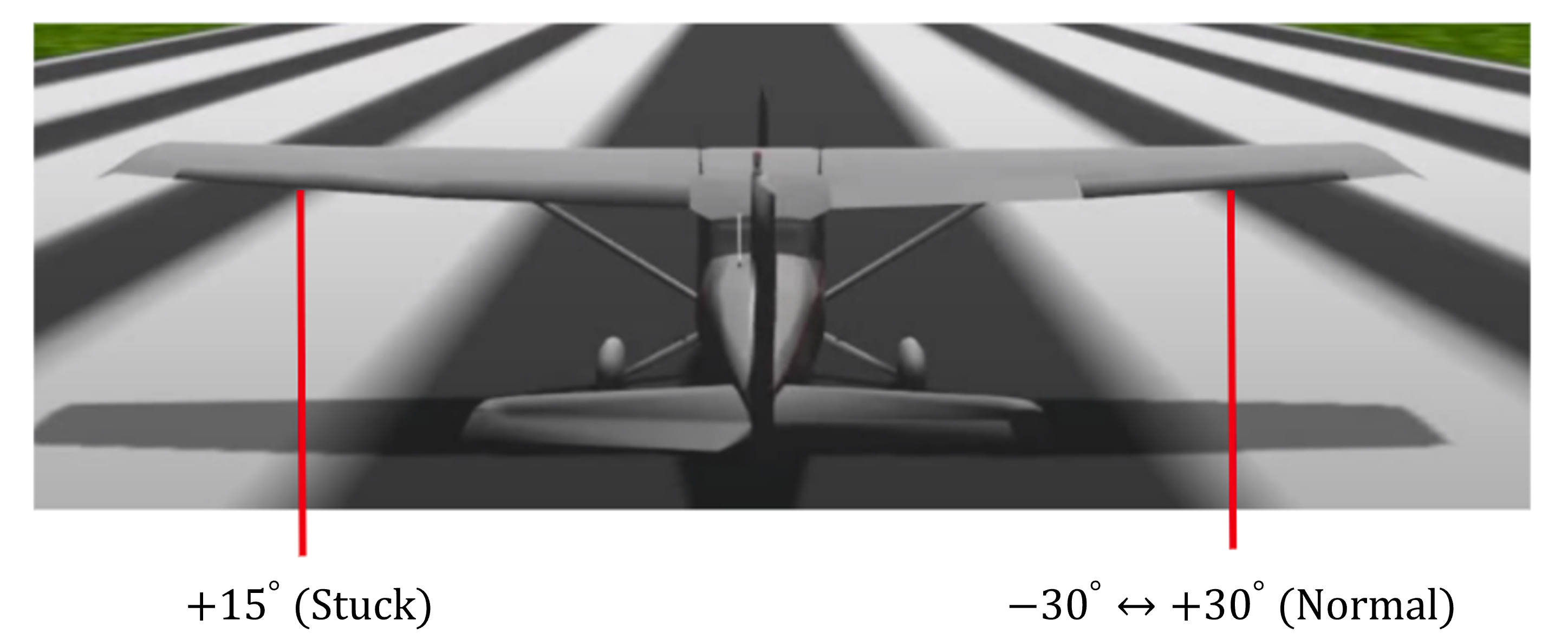}
    \vspace{-1em}
    \caption{
    \textbf{Faulty actuator.}
    The left aileron is stuck at an unknown angle. 
    }
    \vspace{-2em}
    \label{fig:stuck_actuator}
\end{figure}

\begin{figure}[h!]
    \centering
    \includegraphics[width=\columnwidth]{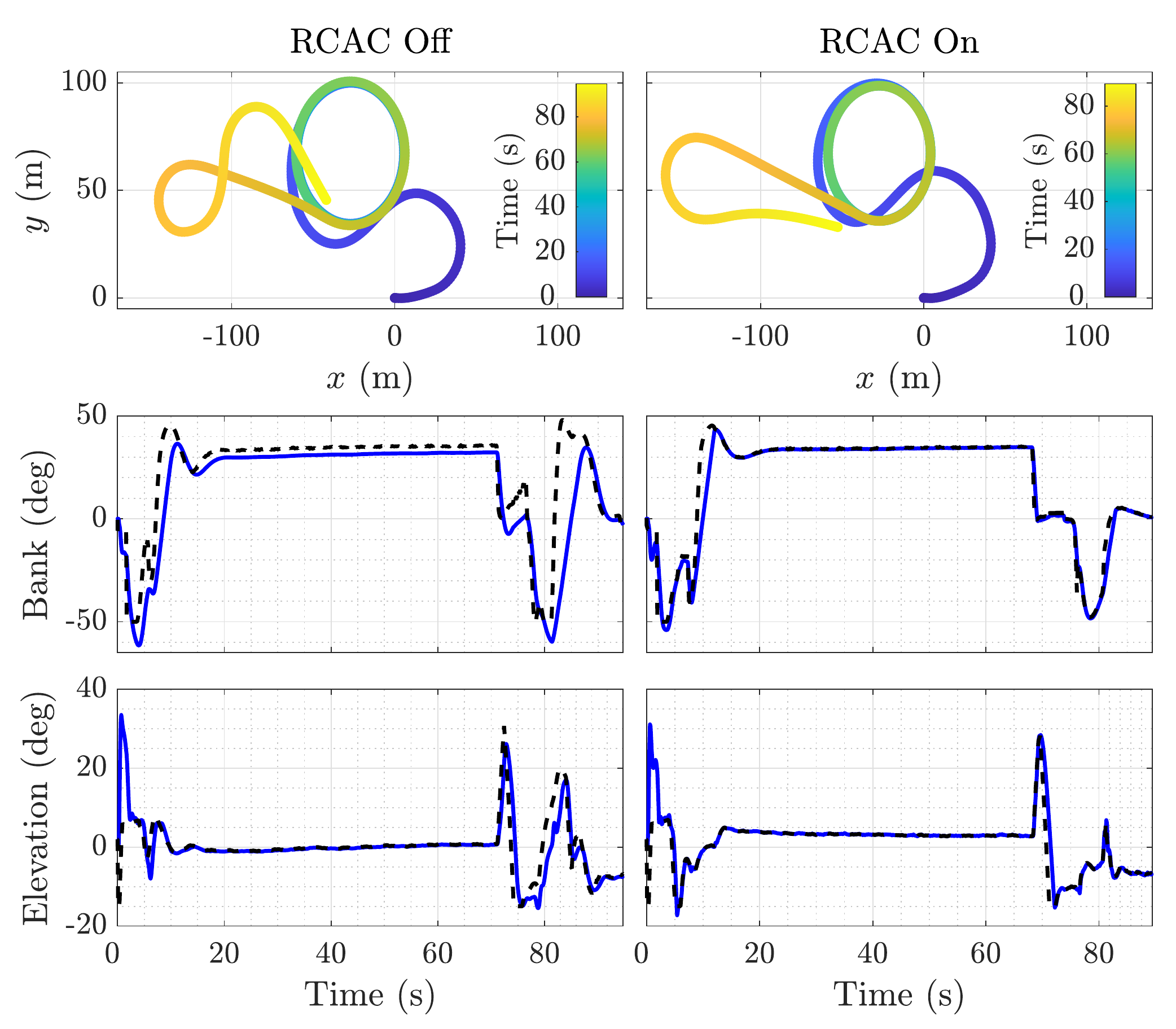}
    \vspace{-1.8em}
    \caption{
    \textbf{Faulty actuator.}
    Ground trace, bank, and elevation response of the aircraft with a faulty actuator.
    The plots on the left and right are obtained with the nominaland adaptive autopilot, respectively.
    Note that, in both autopilots, $\alpha_\rmd=1.$
    }
    \label{fig:Sim_stuck}
\end{figure}

\begin{figure}[h!]
    \centering
    \vspace{1em}
    \includegraphics[width=\columnwidth]{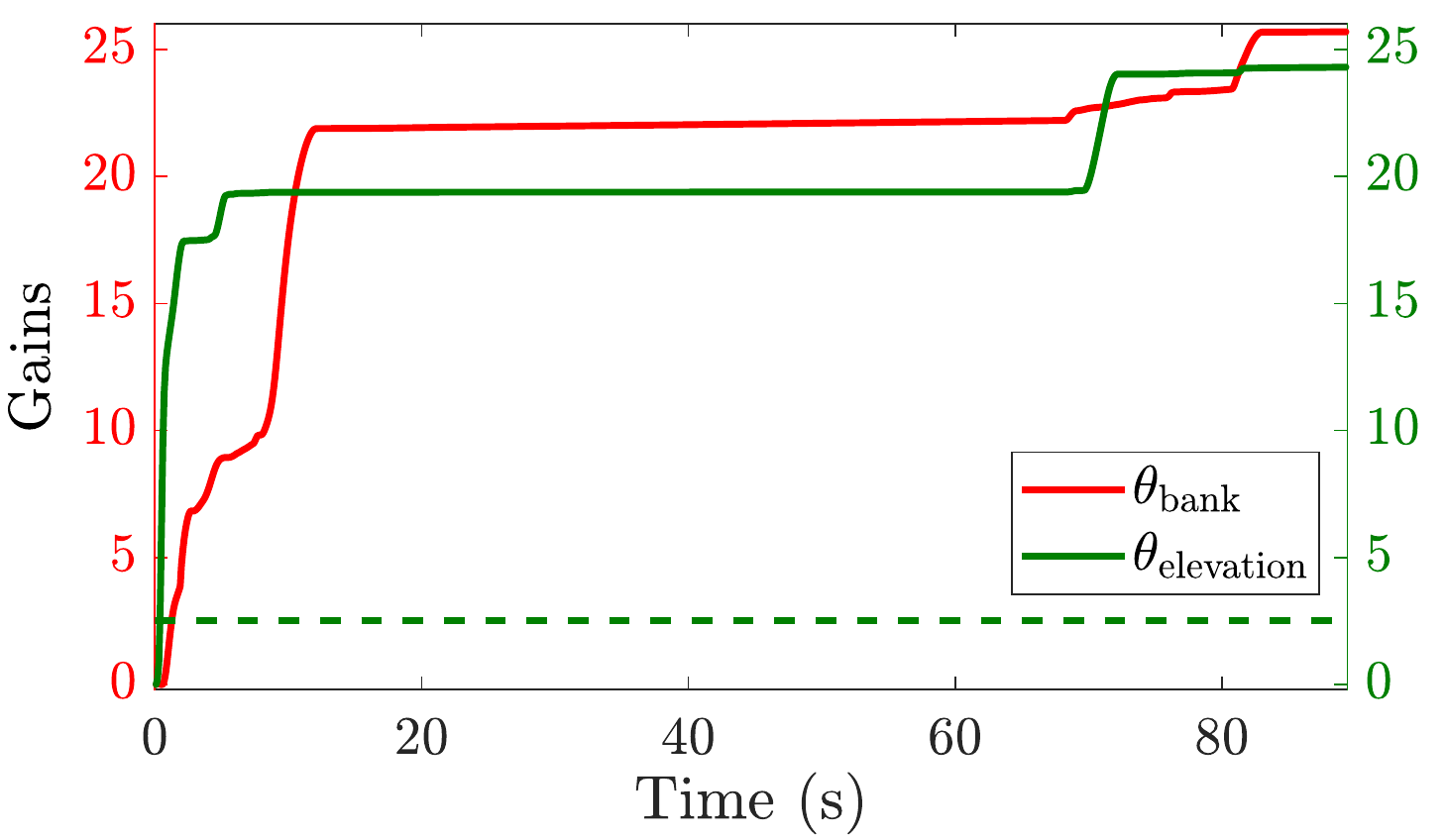}
    \vspace{-2em}
    \caption{
    \textbf{Faulty actuator.}
    Adaptive bank and elevation controller gains optimized by RCAC in the adaptive autopilot for the faulty actuator case.
    The fixed gains are shown in dashes, and the adaptive gains are shown in solid for both the bank and elevator controller. 
    }
    \vspace{-1.5em}
    \label{fig:Sim_RCAC_stuck}
\end{figure}

\begin{figure}[h!]
    \centering
    \includegraphics[width=\columnwidth]{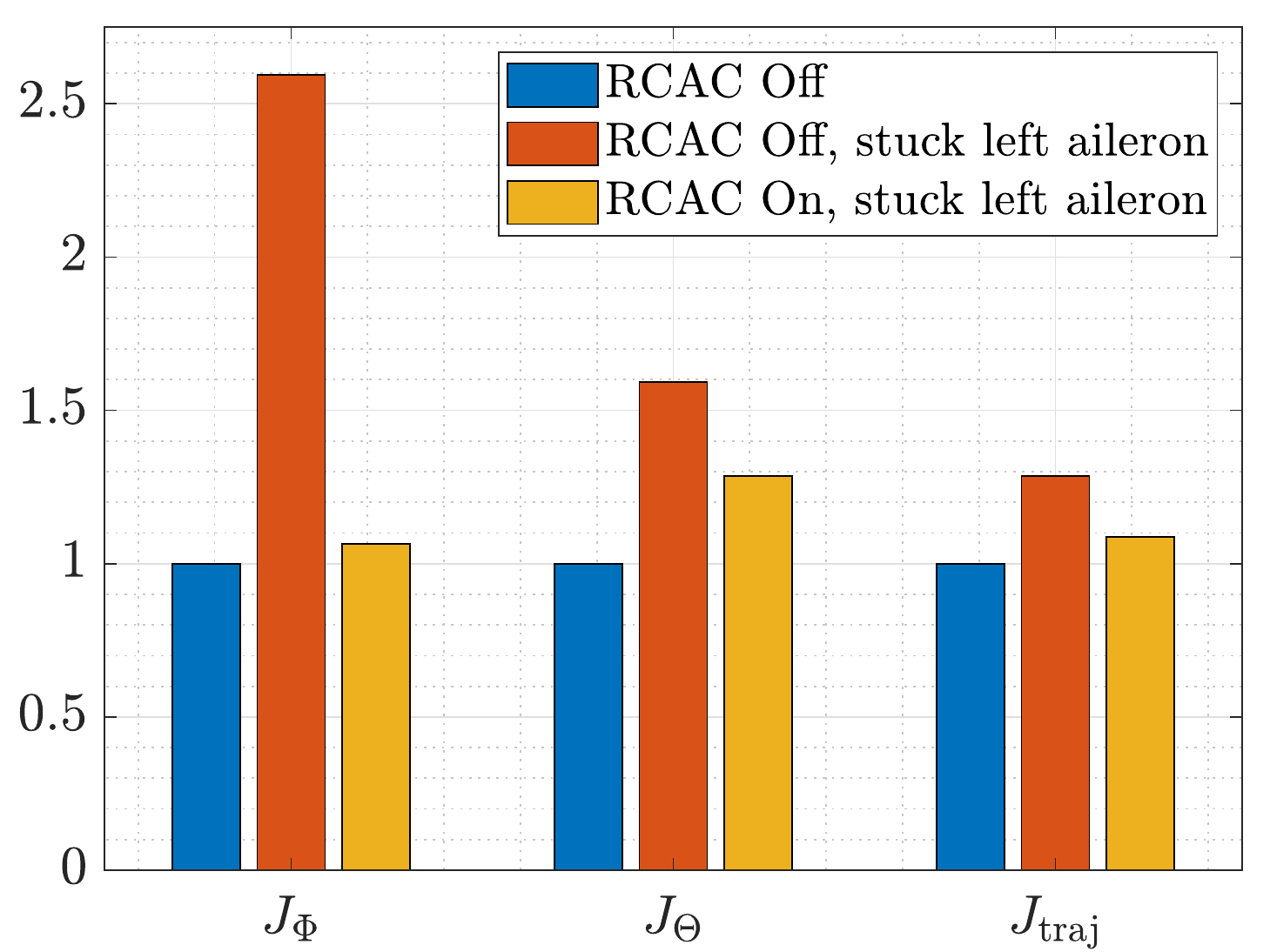}
    \caption{
    \textbf{Faulty actuator.}
    Bank, elevation, and trajectory-tracking error metrics in the case of a faulty actuator with the nominaland adaptive autopilots.
    Note that the error metrics are normalized by the corresponding error metric obtained with the nominal autopilot and healthy actuator.
    }
    \label{fig:Sim_J_cost_stuck}
\end{figure}

Next, we investigate the performance of the adaptive autopilot in the case of faulty actuators.
In particular, we consider the case where an aileron is stuck at an unknown angle as shown in Figure \ref{fig:stuck_actuator}.
With the aileron stuck at an unknown position, we command the aircraft to follow the mission waypoints shown in Figure \ref{fig:FW_sim_trajectory}. 
This test is performed with both the nominal and adaptive controller.
In both cases, note that $\alpha_\rmd = 1.$
Figures \ref{fig:Sim_stuck} and \ref{fig:Sim_RCAC_stuck} show the trajectory-following response in the case where the left aileron is stuck with the nominal and the adaptive autopilot.
Furthermore, Figure \ref{fig:Sim_J_cost_stuck} shows the error metrics in the case of the faulty actuator.
Note that the adaptive autopilot improves the trajectory-tracking error in the case of the faulty actuator and recovers the benchmark performance.

\clearpage

\section{Flight Test Results}
\label{sec:flight_tests_exp}

\begin{figure}[h!]
    \centering
    \includegraphics[width=0.92\columnwidth]{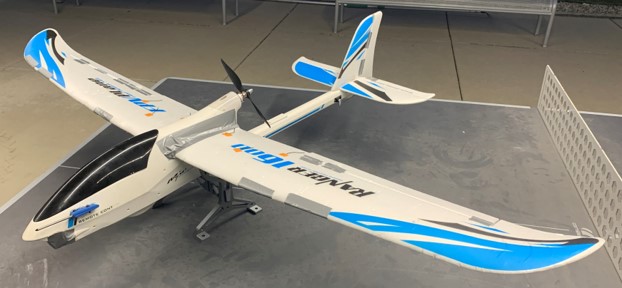}
    \caption{{Volantex Ranger 1600 fixed-wing RC aircraft used in flight experiments.}
    }
    \label{fig:FW_frame}
\end{figure}

This section presents the experimental flight results obtained with the adaptive autopilot. 
%
%
In this work, the flight tests are conducted with a Volantex Ranger 1600 fixed-wing aircraft, shown in Figure \ref{fig:FW_frame}, at the Scio Flyers RC model aircraft club located at $(42.298 \rmN, 83.843\rmW) $. 
To demonstrate the performance improvements due to the adaptive autopilot, the performance of the nominal autopilot is degraded by scaling its fixed gains by the degradation factor $\alpha_\rmd.$
In this work, we focus only on the attitude controller, thus the gains of only the attitunde controller in the nominal autopilot are scaled. 
%
The hyperparameters $P_0,$ $R_u,$ and $\sigma$ used in RCAC are shown in Table \ref{tab:RCPE_variables_Exp}. 
Furthermore, we set $R_z=1$ in all adaptive controllers and all tests. 
Note that once the RCAC hyperparameters are tuned, they are not changed as $\alpha_\rmd$ is varied across the flight tests. 

\begin{table}[h!]
    \vspace{2em}
    \caption{\footnotesize RCAC hyperparameters in the adaptive autopilot for physical flight experiments. }
    \label{tab:RCPE_variables_Exp}
    \centering
    \renewcommand{\arraystretch}{1.2}
    \begin{tabular}{|c|l|l|l|}
        \hline
        \multicolumn{1}{|c|}{\textbf{Controller}}  & \multicolumn{1}{|c|}{${P_0}$}  & \multicolumn{1}{|c|}{${R_u}$} & \multicolumn{1}{|c|}{${\sigma}$}
        \\ \hhline{|=|=|=|=|}
        \eqref{eq:adaptive_elevation_rate_P}, $\theta_\Theta$ & $0.1$ & $0.001$ & $0.1$
        \\ \hline
        \eqref{eq:adaptive_roll_rate_P}, $\theta_\Phi$ & $0.1$ & $0.001$ & $-0.1$
        \\ \hline
    \end{tabular}
\end{table}

\begin{figure}[h]
    \centering
    \includegraphics[width=0.92\columnwidth]{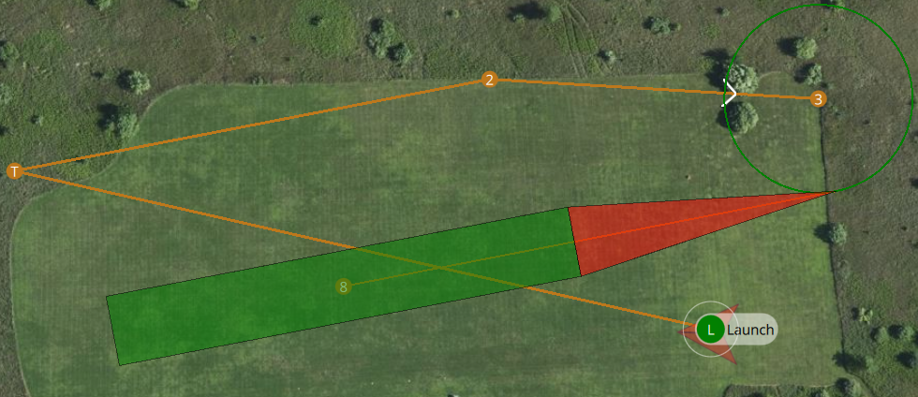}
    \caption{Waypoints used to construct the flight trajectory in physical flight experiments. 
    }
    \label{fig:FW_exp_trajectory}
\end{figure}

The mission waypoints are shown in Figure \ref{fig:FW_exp_trajectory}.
The aircraft is launched by hand from the launch point and is commanded to fly towards point $\rmT$ while climbing to an altitude of 20 m.
The aircraft is then commanded to fly around point 2 in a steady-state circular flight with a radius of 20 m for around 1 to 2 minutes.
Finally, the aircraft is commanded to land along the green strip.
During the takeoff and landing phases, the autopilot is in \textit{stabilized mode}, in which the bank and elevation commands are issued by a pilot. 
During the rest of the flight, the autopilot is in \textit{mission mode}, in which the bank and elevation commands are issued by the outer loop of the autopilot.

%





\begin{figure}[h]
    \centering
    \includegraphics[width=\columnwidth]{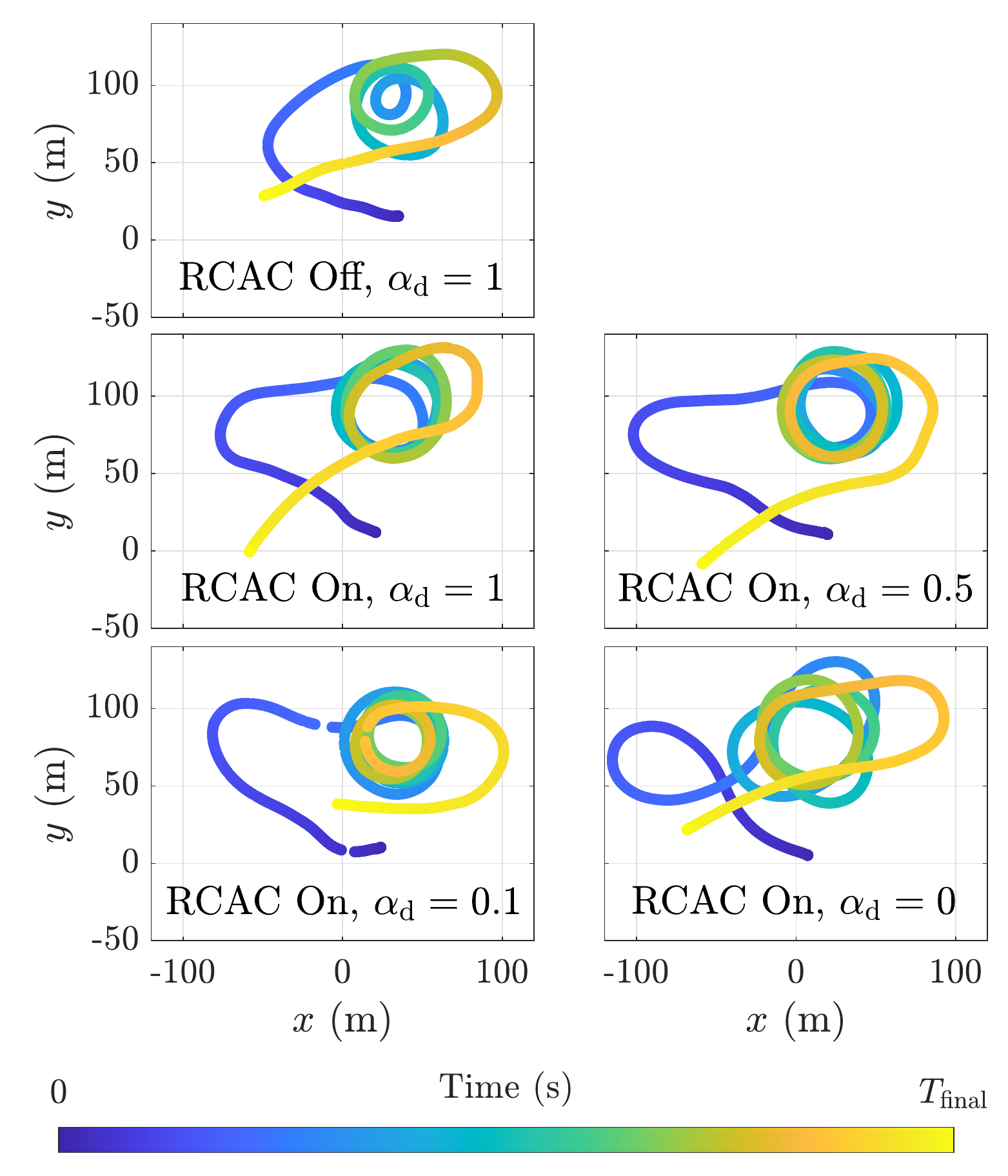}
    \caption{\textbf{Physical flight experiments.}
    Ground trace of the aircraft with the nominaland adaptive autopilot for several values of the degradation factor $\alpha_\rmd.$
    }
    \label{fig:Exp_comp_traj}
\end{figure}

\begin{figure}[h!]
    \centering
    \vspace{0.5em}
    \includegraphics[width=\columnwidth]{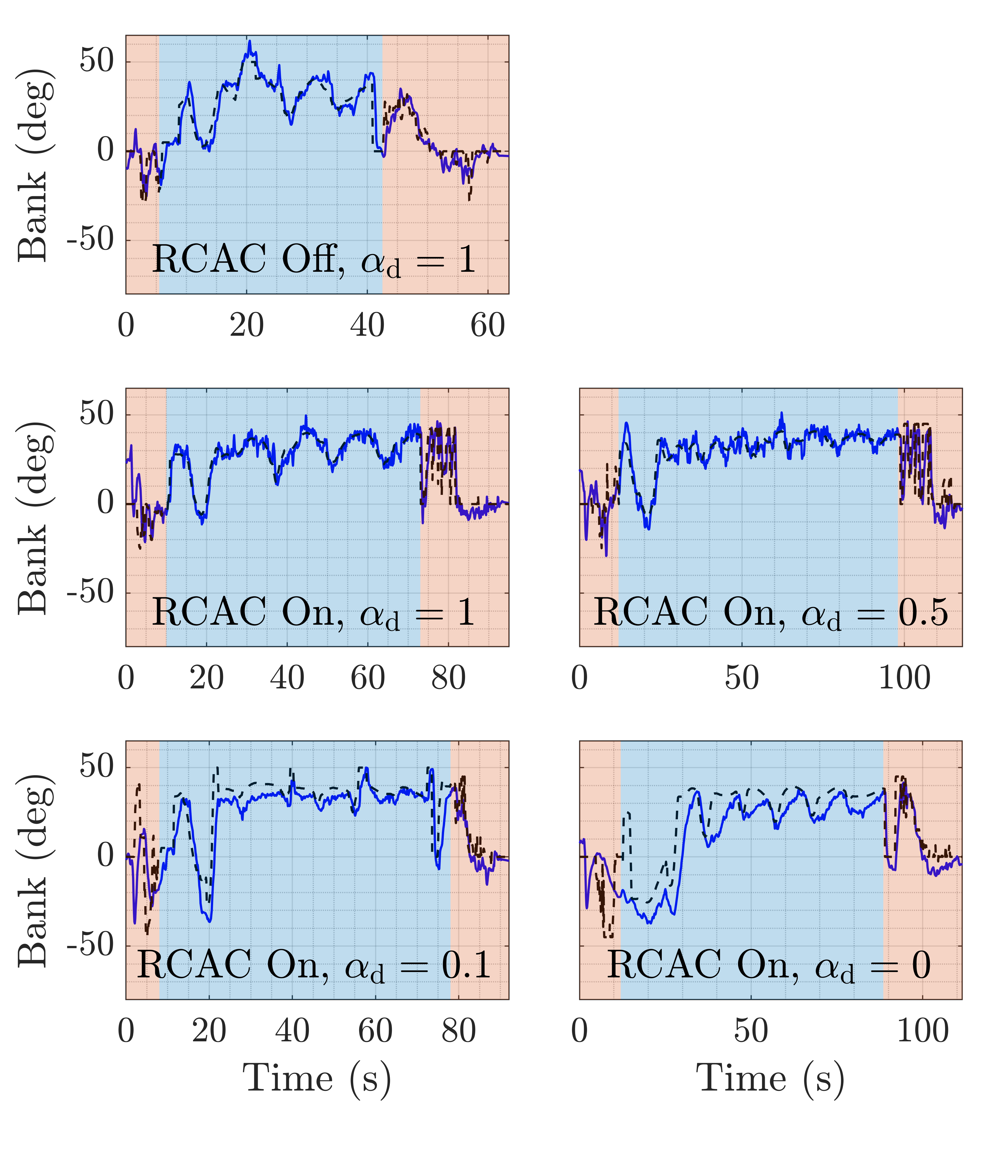}
    \vspace{-3.5em}
    \caption{\textbf{Physical flight experiments.}
    Bank-angle response of the aircraft with the nominaland adaptive autopilot for several values of the degradation factor $\alpha_\rmd.$ 
    The bank angle setpoints are displayed in black dashes.
    Note that the autopilot is in {mission mode} and {stabilized mode} in the blue and orange regions, respectively.
    }
    \label{fig:Exp_comp_bank}
\end{figure}

\begin{figure}[h!]
    \centering
    \includegraphics[width=\columnwidth]{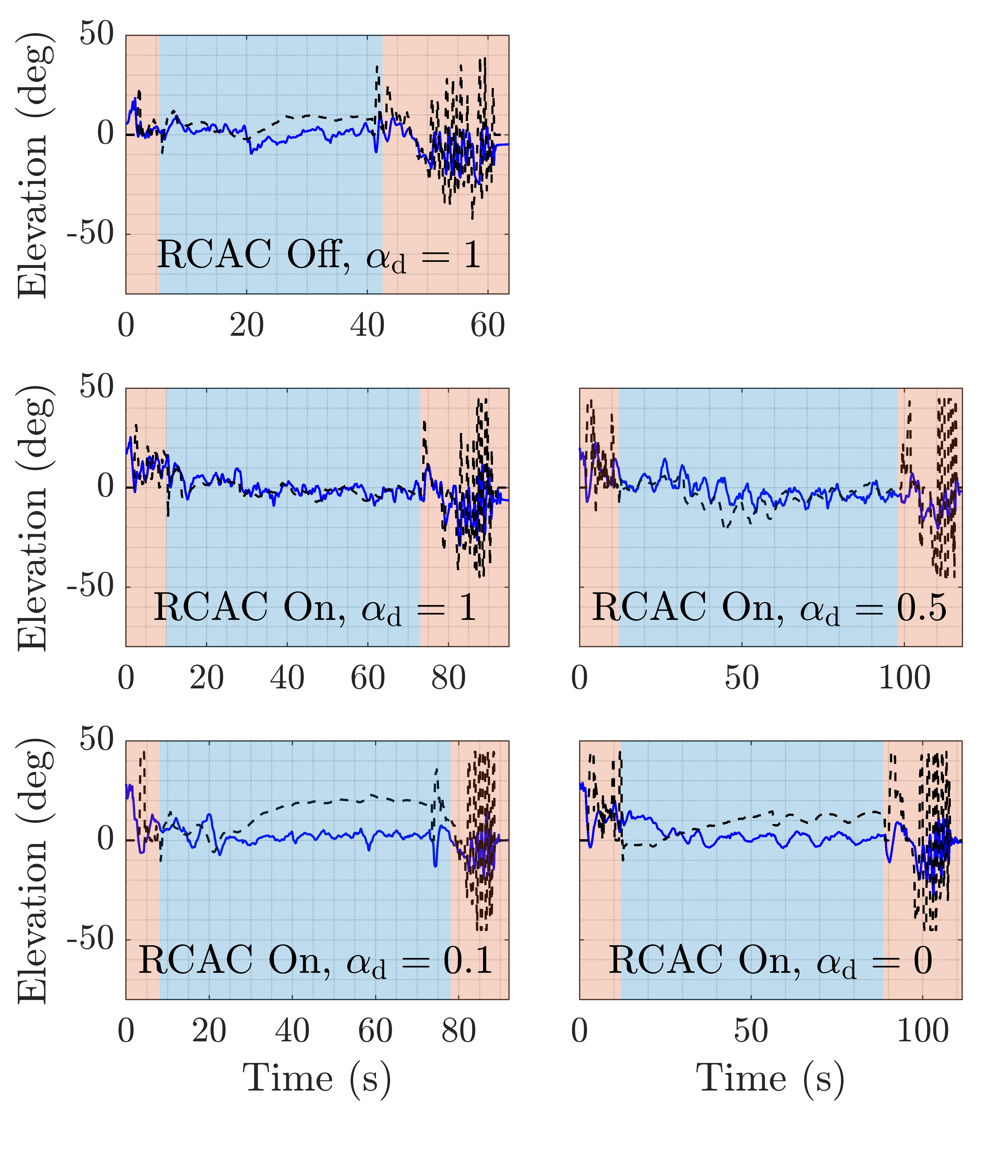}
    \vspace{-3.5em}
   \caption{
    \textbf{Physical flight experiments.}
   Elevation-angle response of the aircraft with the nominaland adaptive autopilot for several values of the degradation factor $\alpha_\rmd.$ 
    The bank angle setpoints are displayed in black dashes.
    Note that the autopilot is in {mission mode} and {stabilized mode} in the blue and orange regions, respectively.
    }
    \label{fig:Exp_comp_elevation}
\end{figure}

\begin{figure}[h!]
    \centering
    \vspace{0.5em}
    \includegraphics[width=\columnwidth]{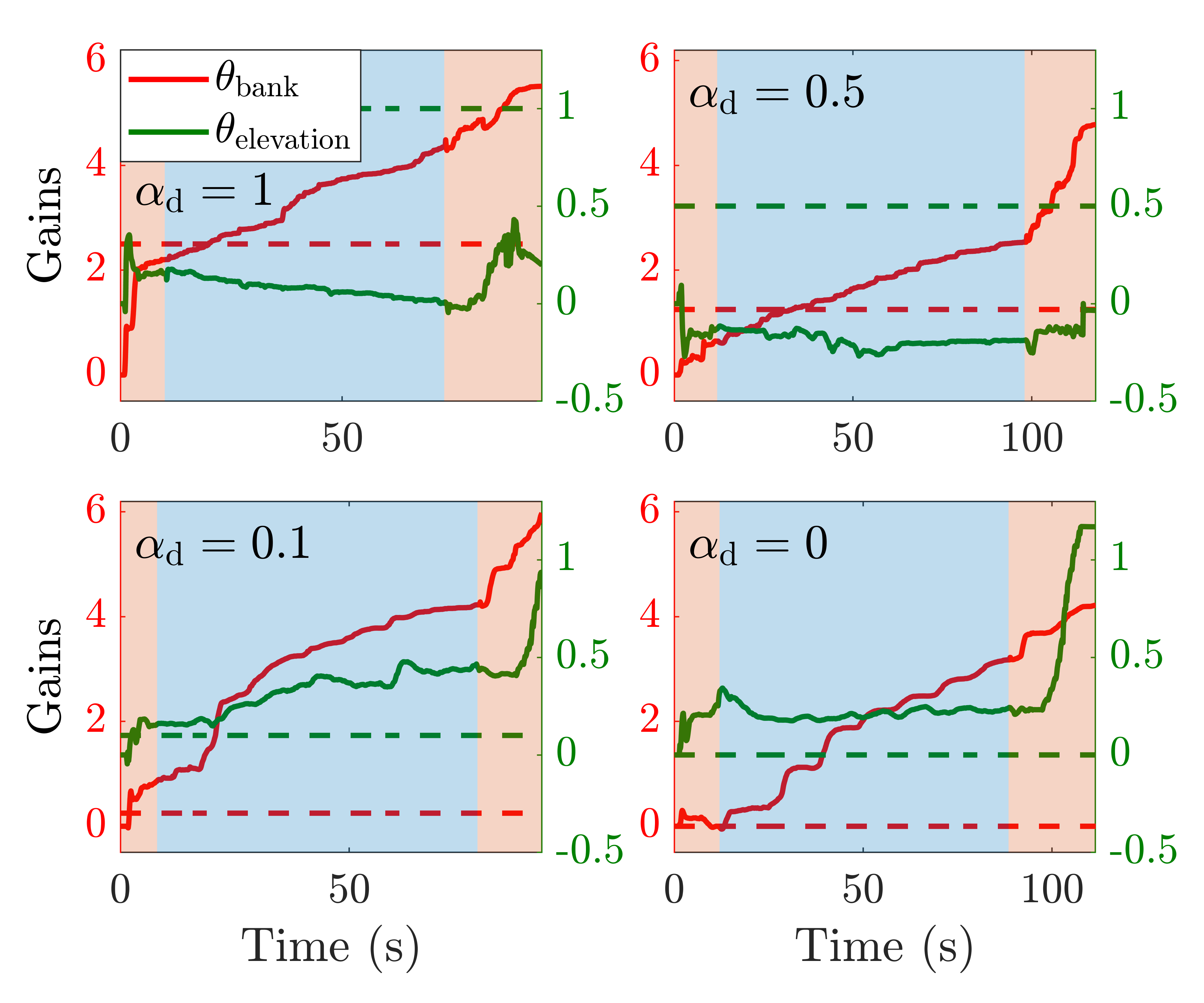}
    \vspace{-2em}
    \caption{
    \textbf{Physical flight experiments.}
    Adaptive bank and elevation controller gains optimized by RCAC in the adaptive autopilot for several values of the degradation factor $\alpha_\rmd.$
    The nominal gains are shown in dashes, and the adaptive gains are shown in solid for both the bank and elevator controller. 
    Note that the autopilot is in {mission mode} and {stabilized mode} in the blue and orange regions, respectively.
    }
    \label{fig:Exp_comp_RCAC}
\end{figure}

\begin{figure}[h!]
    \centering
    \includegraphics[width=\columnwidth]{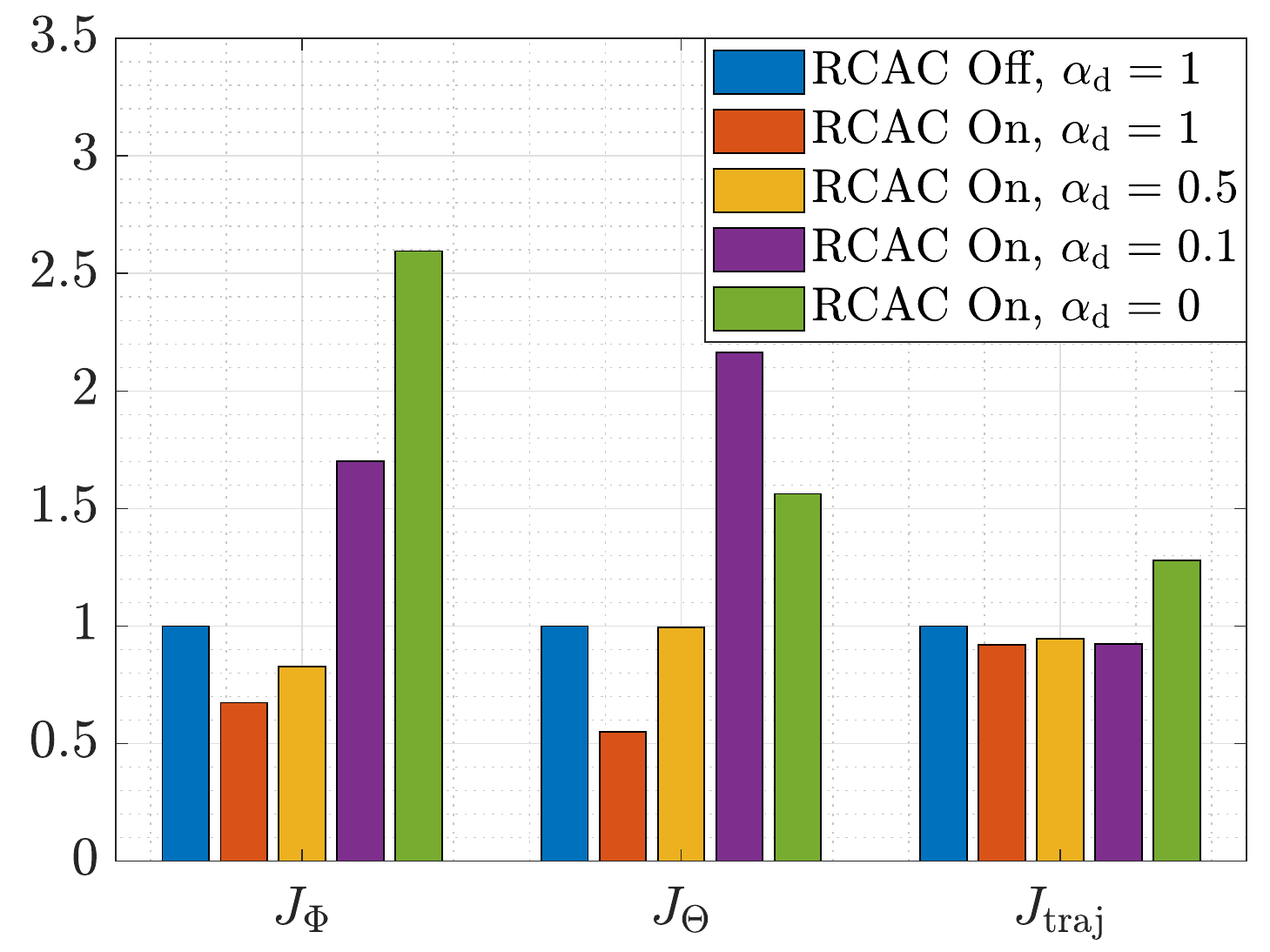}
    \vspace{-2em}
    \caption{
    \textbf{Physical flight experiments.}
    Bank, elevation, and trajectory-tracking error metrics obtained with the adaptive autopilots for several values of $\alpha_\rmd$ and normalized by the corresponding error metric obtained with the nominal autopilot.
    }
    \label{fig:Exp_J_cost}
\end{figure}

Figures \ref{fig:Exp_comp_traj}, \ref{fig:Exp_comp_bank}, and \ref{fig:Exp_comp_elevation} show the ground trace, bank-angle response, and elevation-angle response, respectively, of the aircraft with the nominal and the adaptive autopilot for several values of the degradation factor $\alpha_\rmd.$
Figure \ref{fig:Exp_comp_RCAC} shows the adaptive bank and elevation controller gains optimized by RCAC in the adaptive autopilot for several values of the degradation factor $\alpha_\rmd.$
Figure \ref{fig:Exp_J_cost} shows the normalized error metrics for several values of $\alpha_\rmd$ with the nominal, degraded-nominal, and adaptive autopilot.
The error metrics are normalized by the corresponding metrics obtained with the nominal autopilot, that is, without RCAC and $\alpha_\rmd=1.$
As shown in Figure \ref{fig:Exp_J_cost}, the augmented adaptive autopilot improves the performance over the nominal performance. 
%
%
For $\alpha_\rmd=0.5,$ the trajectory following response degrades substantially with the degraded-nominal autopilot, and in this case, the adaptive autopilot recovers the baseline performance. 
In fact, as the fixed-gain controller is degraded, RCAC compensates by providing larger values of the corresponding gains. 
Finally, the adaptive autopilot is also able to learn the gains from a cold start, that is, the case where the nominal autopilot is completely switched off, that is, $\alpha_\rmd=0.$

\section{Conclusions}
\label{sec:conclusions}

%
This paper presented an adaptive autopilot that can improve an initial poor choice of controller gains in a fixed-gain autopilot, and learn the autopilot gains without any prior knowledge of the dynamics. 
The adaptive autopilot is constructed by augmenting the fixed-gain controllers in an autopilot with adaptive controllers.
%
The adaptive autopilot was used to fly a fixed-wing aircraft model in the Gazebo simulator.
The adaptive autopilot recovered the performance in the case where the fixed-gain autopilot was degraded and learned a set of gains in the case where the fixed-gain autopilot was completely switched off. 
Furthermore, the adaptive autopilot improved the trajectory-tracking performance in the case where the aileron was stuck at an unknown angle in simulation.
The adaptive autopilot was also used to fly fixed-wing aircraft in flight experiments conducted outdoors.
Like the simulation results, the adaptive autopilot improved the flight performance in physical flight experiments. 
%

\vspace{2em}

\renewcommand*{\bibfont}{\small}
\printbibliography

\end{document}